\documentstyle[prb,aps,preprint,eqsecnum]{revtex}
\tightenlines
 
\begin{document}
\widetext
\draft
 
\date{\today }
 
\title{Level Curvature Distribution and the Structure of 
Eigenfunctions in Disordered Systems.}

\author{C.~Basu$^1$, C.M.Canali$^2$\cite{byline}, 
V. E. Kravtsov$^{1,4}$, I. V. Yurkevich$^5$ \\
$^1$International Center for Theoretical Physics, \\
P.O.Box 586, 34100 Trieste, Italy.\\
$^2$ Dept. of Applied Physics,
Chalmers University of Technology\\ and
G\"oteborg University, S-412 96 G\"oteborg, Sweden.\\
$^4$Landau Institute for Theoretical Physics, \\
Kosygina str. 2, 117940 Moscow, Russia.\\
$^5$The University of Birmingham,
School of Physics and Astronomy,\\
Edgbaston, Birmingham B15 2TT, UK.}
\date{today}
\maketitle

\begin{abstract}
The level curvature distribution function is studied both analytically and 
numerically for the case of T-breaking perturbations over the orthogonal
ensemble. The leading correction to the shape of the curvature
distribution beyond the random matrix theory is calculated using the 
nonlinear supersymmetric sigma-model and compared to
numerical simulations on the Anderson model.  

It is predicted analytically and confirmed numerically that the {\it sign}
of the correction is different for T-breaking perturbations caused by a
constant vector-potential equivalent to a 
phase twist in the boundary conditions,
and those caused by a random magnetic field.

In the former case it is shown using a nonperturbative
approach that 
quasi-localized states in weakly disordered systems can cause the
curvature distribution to be nonanalytic. In $2d$ systems the
distribution function $P(K)$ has a branching point at $K=0$ that is 
related to the multifractality of the wave functions and thus should be a 
generic feature of all critical eigenstates. 
A relationship between the branching
power and the multifractality exponent $d_{2}$ is suggested. 
Evidence of the branch cut singularity
is found in numerical simulations 
in $2d$ systems and at the
Anderson transition point in $3d$ systems.
\end{abstract}

\section{Introduction.}
As first suggested  by Edwards and Thouless \cite{Thouless74}, the
sensitivity of the spectrum $\{E_n\}$ of disordered conductors to a small twist
of phase $\phi$ in the boundary conditions $\Psi (x=0,{\bf \rho }
)=e^{i\phi
}\Psi(x=L,{\bf 
{\rho }})$ is a powerful tool to probe the
space
structure of eigenfunctions and distinguish between the extended and the
localized states. More precisely, the quantity $K_n$ that is now referred
to as the ``level curvature'', was introduced in 
Ref.~[\onlinecite{Thouless74}] in
order to describe this sensitivity quantitatively: 
\begin{equation}
K_n=\frac 1\Delta \,\left. \frac{\partial ^2E_n(\phi )}{\partial \phi ^2}%
\right| _{\phi =0},  \label{K}
\end{equation}
where the mean level spacing $\Delta =(\nu L^d)^{-1}$ is related to the
mean density of states $\nu=\langle\nu(E) \rangle $ and the size of the
$d$-dimensional sample $L$.

In complex or disordered quantum systems, the quantity $K_n$ fluctuates over
the ensemble of energy levels $\{E_n\}$ or, for a given level, over the
ensemble of realizations of disorder. The typical width of the distribution
of level curvatures $P(K)$ is of the order of the dimensionless conductance 
$g=D/(L^2\Delta )$, where $D$ is the diffusion coefficient. Thus studying
this distribution one can study the Anderson transition from
metal to insulator that takes place upon increasing the disorder.

Recent work \cite{Alt93} has shown that the distribution of
the fluctuating quantity $K_n$ is a particular example of {\it parametric
level statistics}, i.e. statistics of spectral responses of the system to
a perturbation proportional to some parameter $\phi $.

A remarkable property that the parametric level statistics \cite{Alt93}
share with the usual level statistics \cite{WD,Mehta}, is
that in a certain limit they are universal for all classically chaotic and
disordered systems and can be described by the random matrix theory (RMT) of
Wigner and Dyson \cite{WD,Mehta}. For disordered systems considered
here this limit coincides \cite{Efetov} with $g\rightarrow \infty $. For
chaotic systems the same role is played \cite{Andreev96} by the ratio $%
g=\gamma _1/\Delta $, where $\gamma _1$ is the first nonzero mode in the
spectrum of the Perron-Frobenius operator that describes the chaotic
behavior of the corresponding classical system. In particular, for a
time-reversal-invariant system without spin-dependent interactions
(orthogonal ensemble) the distribution of level curvatures, Eq.(\ref{K}),
was found \cite{ZakrDel93,FvO94,FyodSom95} 
in this limit to have the form: 
\begin{equation}
P_{WD}(k)=\frac 1{2\,(1+k^2)^{\frac 32}},\;\;\;\;k=\frac K{\langle
|K|\rangle },  \label{RMT}
\end{equation}
where $\langle |K|\rangle _{RMT}=2g$ is the average modulus of the level
curvature. Further study \cite{FyodSom95} has shown that the
the form of the curvature distribution is still given by
Eq.(\ref{RMT})
even when weak localization is taken into account;
only the
dimensionless conductance in the expression for $\langle |K|\rangle $ is
changed appropriately.

The form of Eq.(\ref{RMT}) is universal. 
It does not depend, e.g. on the details
of the system and the perturbation. Its validity only requires
the system to be T-invariant with $T^2=1$, and the perturbation to break this
invariance. The underlying physics behind this universality is the
basis invariance of RMT which is equivalent to the 
eigenfunctions of the physical system being structureless.

Anderson localization apparently breaks the basis invariance. Thus 
at a sufficiently small value of $g$ 
the universality of the spectral statistics should break down as well. 
In the strong localization limit one would expect \cite{CanBasu}
a logarithmically normal decay
rather than the power-law tails of the Edwards-Thouless curvature
distribution function.
This is because the fluctuations of level curvature $K\propto e^{-L/\xi}$
can be viewed in this case as the consequence of 
the Gaussian fluctuations of the localization radius $\xi$
for the 
exponentially 
localized wave functions. This picture is qualitatively confirmed by
recent analytical 
\cite{Fyod}
and numerical \cite{Zyczk94,CanBasu} calculations.

In the present paper we present a comprehensive review of our recent 
analytical and numerical
results on the correction to the level curvature distribution 
$\delta P(k)=P(k)-P_{WD}(k)$ when one 
approaches the Anderson transition point $g=g_{d}^{*}$ from the metal side
$g\gg 1$.
Some of the results discussed below are published in 
\cite{CanBasu,b22,KY}. 

It turns out that there are two completely different contributions to the
correction $\delta P(k)$. 
One of them $\delta P_{\rm reg}(k)$ is regular in the small 
parameter $g^{-1}$ and can be obtained by a
perturbative
treatment \cite{KravMir} of the nonzero spatial modes of the 
nonlinear supersymmetric 
sigma-model \cite{Efetov}. 
The main result of this treatment \cite{b22} is that the {\it sign} of the 
correction $\delta P_{\rm reg}(k)$ depends on the
topological nature of perturbation. It is different for the ``global'' 
Edwards-Thouless
curvature, where the perturbation is represented by
a global twist of the phase in the boundary conditions,\cite{ButtImry}
and for the case in which the curvature is probed by a ``local'' 
T-breaking 
perturbation
such as magnetic     
impurities or random magnetic fluxes. 
Below we present a numerical evidence of this fact.

On top of the regular correction, there is \cite{KY} also a
nonperturbative
in $g^{-1}$
correction $\delta P_{s}(k)$ which is proportional to $\exp(-1/g^{-1})$.
The latter correction is due to the so called pre-localized states
\cite{b11,b12,b13,b14,b17}, i.e. eigenstates with anomalously
high peak(s) in the probability density $|\Psi({\bf r})|^2$. 

There 
are reasons to consider the highly irregular, multifractal critical
eigenstates \cite{b9} as the result of a proliferation of such pre-localized
states.
This point of view is partly supported by the observation that
weakly localized states in the critical dimensionality $d=2$
also exhibit a (weak) multifractality, as can be shown
by means of the same methods
(renormalization group \cite{b10,b11} or space
inhomogeneous
saddle-point approximation \cite{b12}) that were used
to discover the pre-localized states responsible for the slow current
relaxation in disordered conductors\cite{b10,b14}.  

This idea enables us to extend the results for $\delta P_{s}(k)$ obtained by
the novel saddle-point
approximation \cite{b14,KY} for $2d$ metals
to the critical state at the Anderson transition in $2+\epsilon$
dimensions. 
Thus we can explain the branching nonanalyticity at $k=0$
found numerically in \cite{CanBasu} for the $3d$
critical level curvature distribution function $P_{c}(k)$. 
Furthermore, we suggest a relationship between the branching
power and the exponent $d_{2}$ describing  generic multifractal critical
states \cite{b9}. This relationship fits well the numerical results and
provides
a link between the spectral statistics and statistics of wavefunctions
near the Anderson transition.

The paper is organized as follows. In Section II we describe the
perturbative in $g^{-1}$ approach for calculating $\delta P_{\rm reg}(k)$.
In Section III we generalize the instanton approximation of 
Ref.~[\onlinecite{b14}]
for the problem of level curvature distribution and calculate the
nonperturbative contribution $\delta P_{s}(k)$ for the metallic 
(weakly-localized) states in quasi-$1d$ and $2d$ systems. In Section IV
we extend the results of Section III to the critical states in
$2+\epsilon$
dimensions and derive the relationship between the branching
nonanalyticity in $P_{c}(k)$ and the fractal dimensionality $d_{2}$.
In Section V the results of numerical simulation on the Anderson model 
are presented. Some open questions are discussed in the Conclusions.

\section{Perturbative corrections to $P(k)$.}
\subsection{Main results.}
\label{mres}

A general approach to calculate the $1/g$-corrections
using the nonlinear supersymmetric sigma-model \cite{Efetov} has
been suggested in \cite{KravMir} and applied to the distributions of different
quantities \cite{FyodSom95,KravMir,FyodMir95}. It is based on
a perturbative analysis of the nonzero 
diffusion modes which are
integrated out to produce $1/g$ corrections to the zero-mode supersymmetric
sigma-model \cite{Efetov}. The latter must then be handled exactly. 

Before going into the details of the calculations we would like to formulate 
the main results for $\delta P_{\rm reg}(k)$ for the case of T-breaking
perturbations over the orthogonal ensemble: 
\begin{equation}
\delta
P_{\rm reg}(k)=C_{d}\;\frac{2-11k^2+2k^4}{2\,(1+k^2)^{7/2}},\;\;\;\;\;k=\frac
K{\langle |K|\rangle }\ll g,  
\label{MR}
\end{equation}
where 
\begin{equation}
C_{d}=\frac 1{(\pi g)^2}\sum_{{\bf q}\neq 0}\frac 1{({\bf q}^2)^2}\times
\left\{ 
\begin{array}{ll}
\left( \frac 4d-1\right) , & {\rm case\ I\ } \\ 
-1, & {\rm case\ II\ }
\end{array}
\right. .  \label{C}
\end{equation}
Here ${\bf q}=\{q_1,...q_d\}$, where $q_i=2\pi n_i$, ($n_i=0,\pm 1,\pm 2...$%
) in the case of the periodic boundary conditions (for an unperturbed
system) considered in this paper.

A remarkable feature of Eqs.(\ref{MR}),(\ref{C}) for $d<4$ is that the {\it %
sign} of the correction is different for global ({\bf case I}) and local
({\bf case II})
T-breaking perturbations. See Fig.~\ref{fig1}. 

The positive sign at small $k$ in {\bf case I} reflects
the tendency towards a weaker spectral response to a change in the boundary
conditions with decreasing $g$. 
It is related to long-range correlations 
in the wavefunctions that result in mesoscopic fluctuations
of the matrix element of perturbation. The same long-range correlations
cause mesoscopic fluctuations of the diffusion coefficient.

The negative sign of the correction
$\delta P_{reg}(k)$ for local perturbations [{\bf case II}] is
entirely due to the effect of the energy level statistics that lead
to mesoscopic fluctuations of
the density of states. The mesoscopic fluctuations of the matrix
elements are suppressed in this case, because the effect of long-range
correlations of the unperturbed wavefunctions is cut by the local nature
of the perturbation.

In order to illustrate the effect of the energy level statistics on the level
curvature we invoke the expression for $K_{n}$ in terms of the matrix
element of the perturbation $|V_{nm}|^{2}$ and the exact eigenvalues in the
absence of perturbation $E_{n}$:
\begin{equation}
\label{sopt}
K_{n}= 2\sum_{m\neq n}\frac{|V_{nm}|^{2}}{E_{n}-E_{m}}.
\end{equation}
From this expression we can clearly identify the two sources of 
the fluctuations in the level curvature.
If the fluctuations of the matrix elements are suppressed,
$|V_{nm}|^{2}$ can be replaced by a constant.
Then there is only one source of fluctuations left, that is the energy
level statistics. Notice that {\it implicitly} this contains 
the effect of the statistics of the {\it eigenfunctions} as well, and
in particular, their long-range correlations. Upon decreasing $g$
the system of energy levels becomes less and less correlated.
In the extreme limit of uncorrelated levels the distribution of curvatures
is known \cite{Thouless74} to become of the Cauchy-Lorentz form
$P_{CL}(k)=1/\pi (1+k^2)$. Since $P_{CL}(0)=1/\pi$ and $P_{WD}(0)=1/2$
we conclude that the effect of softening the energy level correlations on
the shape of the curvature
distribution is such that $\delta P(0)<0$, in
full agreement with
the results for the {\bf case II}.

The principal result of this Section is that Thouless
relationship of proportionality $\langle|K|\rangle \propto g$ breaks down
beyond RMT.
The ratio
$r(g)=\langle|K|\rangle /2g$ {\it increases} above its RMT value $r=1$,
the correction being equal to:
\begin{equation}
\label{r}
\delta r(g)=\frac{\delta\langle|K|\rangle}{2g}=
\frac 1{(\pi g)^2}\sum_{{\bf q}\neq 0}\frac 1{({\bf q}^2)^2}\times
\left\{
\begin{array}{ll}
\left( \frac{9}{2}-\frac{16}{d}+\frac{36}{d(d+2)}\right) , & {\rm case\ I\
} \\
\hspace {1cm}\frac{9}{2}, & {\rm case\ II\ }
\end{array}
\right. . 
\end{equation}

\subsection{Functional representation for $P(K)$.}

We know describe how Eqs.~(\ref{MR}),(\ref{C}) have been obtained.
Let us consider a disordered mesoscopic $d$-dimensional system with a
random white-noise impurity potential $V({\bf r})$ perturbed by a small
vector-potential $\overrightarrow{\phi}/L$. It is described by the
microscopic Hamiltonian of the form:
\begin{equation}
H=\frac 1{2m}\left( \frac 1i\nabla -\frac{\overrightarrow{\phi }}L\right)
^2+V\left( {\bf r}\right) ,
\end{equation}

In {\bf case I} we assume the 
sample to be closed into a ring geometry
pierced by a small static magnetic flux
$\Phi$. Then the effect of the perturbation is equivalent to a twist 
of the boundary conditions generating the phase $\phi=2\pi
\Phi/\Phi_{0}$ ($\Phi
_0=hc/e$ is the flux quantum).  As a vector potential
we will take a constant vector of the form
$\overrightarrow{\phi }/L=(\phi/L)\,{\bf n}=(\phi/L)\{1,0,..0 \}$.

In the ``local'' {\bf case II} we consider
$\overrightarrow{\phi}({\bf r})$ to
be a random $\delta$-correlated vector-potential: 
\begin{equation}
\label{vp}
\langle \phi_{\alpha}({\bf
r})\,\phi_{\beta}({\bf r'}) \rangle = v_{\tau}\phi^2 \,\delta({\bf r}-{\bf
r'})\,\delta_{\alpha \beta},\;\;\;\; v_{\tau}=\frac{D}{2\pi\nu v_{F}^2}.
\end{equation}
The parameter $\phi$ is introduced in this way in order to keep
$\langle|K|\rangle_{RMT}=2g$ the same as to the ``global'' curvature
(case I). 

The curvature distribution function 
\begin{equation}
P\left( K\right) =\Delta \left\langle \sum_n\delta \left( K-K_n\right)
\delta \left( E-E_n\right) \right\rangle
\end{equation}
can be expressed in terms the two-level parametric correlation function $%
R\left( \omega ,\phi \right) $ 
\begin{equation}
\label{TLCF}
R\left( \omega ,\phi \right) =\nu^{-2}\left\langle \nu \left( E+\omega
,\phi \right) \nu \left( E,\phi =0\right) \right\rangle 
\end{equation}
in a form similar to one derived in Ref.~[\onlinecite{KravZirn}] for the
distribution of level velocities: 
\begin{equation}
\label{P-R}
P\left( K\right) =\lim_{\phi \rightarrow 0}\frac{\phi ^2}2R\left( \omega
=\frac{1}{2}K\Delta\phi^2 ,\phi \right) .
\end{equation}
Indeed, using the exact expression for the fluctuating density of states
$\nu(E,\phi)=L^{-d}\sum_{n}\delta(E-E_{n}(\phi))$ we have:
\begin{equation}
\label{KZ}
R(\omega,\phi)=\Delta^2 \sum_{m,n}\langle\delta(\omega
+E_{n}(0)-E_{m}(\phi))\,\delta(E-E_{n}(0))\rangle.
\end{equation}
Because of the level repulsion, in the limit $\omega\rightarrow 0$ and
$\phi\rightarrow 0$ only terms
with $n=m$ contribute to the
sum Eq.(\ref{KZ}). On the other hand, with a perturbation
which is odd under time reversal, the T-invariant level energy 
 $E_{n}(\phi)\approx
E_{n}(0)+\frac{1}{2}K_{n}\Delta\,\phi^2$ must be even\cite{note1}
in $\phi$. Then choosing $\omega=\frac{1}{2}K\Delta\,\phi^2$ we
immediately
arrive at Eq.(\ref{P-R}).

The two-level correlation function $R(\omega,\phi)$ can be represented
in the form of a functional integral using the 
Efetov's supersymmetry approach. A straightforward application of 
the results of Ref.~[\onlinecite{Efetov}] and Eq.(\ref{P-R}) leads to: 
\begin{equation}
P\left( K\right) =-\frac{1}{64}\left. \frac{\partial ^2Z}{\partial
J_1\partial J_2}%
\right| _{J_1=J_2=0},\quad Z=\lim_{\phi \rightarrow 0}\left\{\phi^2\,
\Re\int DQ\exp \left( -F\left[ Q\right] \right)\right\} ,  \label{main}
\end{equation}
where for the {\bf case I} the functional $F[Q]$ takes the form:
\begin{equation}
F\left[ Q\right] =-\frac{\pi g}{8}\int \frac{d{\bf r}}VStr\left\{
\left(L \nabla Q+i\left[ \widehat{\phi },Q\right] \right) ^2+2ik
\phi ^2 (\Lambda Q)\right\}+\int \frac{d{\bf r}}VStr(JQ) ,  \label{F}
\end{equation}
A similar representation for $P(K)$ has been used in 
Ref.~[\onlinecite{FyodSom95}].

In Eq.(\ref{F}) we have introduced the notation \cite{note2}

\[ $k=K/(2g)$,\] 
\[
\widehat{\phi }=\overrightarrow{\phi }P_{+} \tau ,\quad \quad J=%
\widehat{k}\left[ J_1P_{+} +J_2P_{-}\right],
\]
and
\[
P_{+} =\frac 12\left( 1+\Lambda \right) ,\qquad P_{-}
=\frac 12\left( 1-\Lambda \right).
\] 

The coordinate dependent $8\times 8$ supermatrices $Q\left( {\bf r}\right) $
are parametrized as $Q=T^{-1}\Lambda T$, where $T$ belongs to a graded coset
space $UOSP\left( 2,2|4\right) /\\UOSP\left( 2|2\right) \otimes UOSP\left(
2|2\right) $ \cite{VWZ85}. 

Other matrices are specified as follows: 
\begin{eqnarray*}
\Lambda &=&diag\left( I_2,I_2,-I_2,-I_2\right) _{R-A},\qquad \tau
=diag\left( \tau _3,\tau _3,0,0\right) _{R-A}, \\
\widehat{k} &=&diag\left( I_2,-I_2,I_2,-I_2\right) _{R-A},\qquad \tau
_3=diag\left( 1,-1\right) ,\quad I_2=diag\left( 1,1\right) .
\end{eqnarray*}
Above we imply the following hierarchy of blocks of supermatrices:
retarded-advanced $\left( R-A\right) $ blocks, boson-fermion $\left(
B-F\right) $ blocks, and blocks corresponding to time reversal.
\\In the {\bf case II} the linear in $\phi$
term in Eq.(\ref{F} ) is absent but otherwise the functional $F[Q]$
is the same provided that $\phi$ is introduced as in Eq.(\ref{vp}). 
A similar functional $F[Q]$ appears \cite{Efetov} if one considers a small
concentration of magnetic impurities as a perturbation. In both cases the
structure of the ``covariant derivative'' ${\cal D}Q= \nabla Q+\frac
iL\left[ \widehat{\phi },Q\right]$, which implies a sort of global gauge
invariance, is broken down. 

It is important in deriving the functional $F[Q]$
for the case II that the correlation radius of the random vector-potential
is much smaller than the elastic scattering length. In this case the
averaging over $\overrightarrow{\phi}({\bf r})$ should be done {\it before}
switching to $Q$-variables that are assumed to be
slowly varying in space.
In the opposite
limit of large correlation radius, one should average $e^{-F[Q]}$ over
$\overrightarrow{\phi}({\bf r})$ and arrive at a much more complicated
functional.

\subsection{Perturbative treatment of nonzero modes.}

The representation given in Eqs.(\ref{main})-(\ref{F}), 
in terms of the field $Q({\bf r})$,
contains all the spatial diffusion modes $\gamma _q=(D/L^2){\bf q}^2$.
However, in doing the limit $\phi \rightarrow 0$ in Eq.(\ref{main}) the
main role is played by the zero mode that corresponds to ${\bf q}=0$. At
$\phi =0$ this mode is gapless and thus it does not cost any energy 
no matter how large are the
components of the field $Q$ in the noncompact boson-boson sector \cite
{Efetov}. It is the arbitrarily large amplitudes of the zero mode components
of the field $Q$ that compensate the infinitesimal parameter $\phi $ in Eqs.(%
\ref{main})-(\ref{F}) and lead to a finite result for $P(K)$. Thus the space
independent zero mode $Q_0$ must be considered nonperturbatively. 

In the
limit $g\rightarrow \infty $ all the nonzero modes can be neglected \cite
{Efetov}, and one arrives \cite{FyodSom95} at the RMT result, Eq.(\ref{RMT}).
For finite $1/g$ the nonzero modes should be also taken into account.
However, all the nonzero modes can be treated perturbatively for $g\gg 1$
leading to some corrections to the zero-mode action. In order to obtain
these corrections we have to separate the zero modes from all other modes and
then integrate over all the nonzero modes using a certain perturbative
scheme.

Following the method suggested by
Kravtsov and Mirlin \cite{KravMir} we decompose matrices $Q\left( {\bf r}%
\right) $ as follows: 
\begin{equation}
Q\left( {\bf r}\right) =T_0^{-1}\widetilde{Q}\left( {\bf r}\right) T_0,\quad 
\widetilde{Q}=\Lambda \frac{1+\widetilde{W}/2}{1-\widetilde{W}/2}
\label{separ}
\end{equation}
where $T_{0}$ describes the zero mode and $\widetilde{W}({\bf
r})=\sum_{{\bf
q}\neq
0} \widetilde{W}_{{\bf q}}\,e^{i{\bf q}{\bf r}}$ does not contain the zero
mode at all.

As has been already noticed, the main contribution to the functional
integral comes from the zero mode. The zero-mode approximation,
$Q=Q_{0}=T_{0}^{-1}\Lambda T_{0}$ is known to be equivalent to 
RMT\cite{VWZ85}. To go beyond RMT we integrate
perturbatively over $\widetilde{W}({\bf r})$ to obtain the effective
zero-mode
action $F^{eff}\left[ Q_0\right] $ as follows: 
\[
F^{eff}\left[ Q_0\right] =-\ln \int D\widetilde{Q}\cdot J\left[ \widetilde{Q}%
\right] \exp \left\{ -F\left[ Q_0,\widetilde{Q}\right] \right\}, 
\]
where $F\left[ Q_0,\widetilde{Q}\right]$ is obtained from $F[Q]$ by 
substituting the decomposition of Eq.(\ref{separ}) and $J[\tilde{Q}]$
is the Jacobian of the corresponding nonlinear transformation.

This scheme is implemented in Appendix \ref{app1}. 
As a result we have the effective
action expanded up to the second order in $1/g$: 
\begin{equation}
F^{eff}\approx F_0^{eff}+F_1^{eff}+F_2^{eff}+F_J^{eff}.  \label{eff}
\end{equation}
The first term in Eq.(\ref{eff}) $F_0^{eff}\left[ Q_0\right] $ is nothing but
the zero-mode action responsible for the $RMT$-results\cite{note3}:
\begin{equation}
\label{0}
F_0^{eff}=\frac 12\alpha -ik\beta-\sum_{p}J_{p}\sigma_{p} ,
\end{equation}
where $p=\pm$ labels the $(R-A)$ blocks and 
\begin{equation}
\alpha
=STr\left( \hat{\phi} Q_{0} \right)
^2,\quad \beta =STr\left(\hat{\phi}^2 Q_{0}\right),\quad
\sigma_{p}=STr[\widehat{k}Q_{0}^{pp}].
\end{equation}
The next term $F_1^{eff}\left[ Q_0\right] $ is the first order (weak
localization) correction obtained by Fyodorov and Sommers \cite{FyodSom95}: 
\begin{equation}
F_1^{eff}=-\frac 12\Pi _2\cdot \alpha ,
\end{equation}
where
\begin{equation}
\qquad \Pi_{2n}=\sum_{{\bf
q}}\Pi^{n}
\left( {\bf q}\right) \equiv \sum_{{\bf q}}\frac 1{(\pi g\, {\bf q}^2)^n}.
\end{equation}
The first order correction $F_1^{eff}$ leads to a renormalization of the
coefficient
in front of $\alpha$ in $F_0^{eff}$ and can be absorbed in the
dimensionless conductance $g$. It can be checked that the renormalized
coefficient $\bar{g}$ is exactly the conductance with weak localization
corrections taken into account. In what follows we shall assume $g$ to be a
renormalized conductance and we shall omit $F_1^{eff}$.

We now consider the higher-order term $F_2^{eff}\left[
Q_0\right] $: 
\begin{eqnarray}
\label{f2}
F_2^{eff}&=&\Pi _4
\left\{\left(-\frac{3}{4}+\frac{4}{d}-\frac{9}{d(d+2)}
\right)\,\alpha^2 +\left(\frac{7}{2}-\frac{16}{d}+\frac{36}{d(d+2)}
\right)\,\beta^2 \right\} \\ \nonumber
&+& \frac{1}{2}\Pi_{4}\,\left(k^2\beta^2 +ik(1-4/d)\,\alpha\beta+\alpha 
\right),
\end{eqnarray}
and a higher-order source-induced contribution ${\cal F}_J\left[
Q_0\right] $: 
\begin{equation}
\label{fj}
F_J^{eff}=
\Pi_4\cdot
\left\{ \left(
\frac{1}{2}(1-4/d)\,\alpha-ik\,\beta\right)
\sum_pJ_{p}\,\sigma_{p}
-\prod_pJ_p\,\sigma_{p}\right\} . 
\end{equation}
The terms in $F^{eff}$ containing the factors $1/d$ and $1/d(d+2)$
originate, after the angular integration over ${\bf q}$,
from the {\it gradient} term 
linear in $\overrightarrow \phi$, 
$Str[\overrightarrow \phi \nabla Q\,Q]$, in
Eq.(\ref{F}). Such term is present only in the ``global'' {\bf case I}.
For this reason in the ``local'' {\bf case II} all the $d$-dependent 
terms in $F^{eff}$ should be omitted.

Differentiating the partition function $Z$ with respect to the sources
we arrive at the expression for the
level curvature distribution function $P\left( k\right)$, where
$k=K/(2\bar{g})$: 
\[
P\left( k\right) =\lim_{\phi \rightarrow 0}\frac{-\phi ^2}{32\pi \bar{g} 
}%
\Re\int DQ_{0}\cdot \Phi \left( \alpha ,\beta \right) \cdot \exp \left[
-\frac
\alpha 2+ik\beta \right] Str\widehat{k}Q_{0}^{11}Str\widehat{k}Q_{0}^{22}, 
\]
\begin{equation}
\Phi \left( \alpha ,\beta \right) =1-F_2^{eff}+\Pi _4\left( 1+2ik\beta 
-(1-4/d)\,\alpha\right) .  
\label{s}
\end{equation}
A remarkable feature of the perturbation theory is that the function
$\Phi \left( \alpha ,\beta \right) $ is a polinomial in $%
\alpha $ and $\beta $. 
This property is due to the fact that long-trace vertices that appear
after the perturbative integration over $\tilde{W}({\bf r})$ factorize
into the product of short-trace vertices $\alpha$,$\beta$, and
$\sigma_{p}$.

Then the correction $\delta P(K)$ can be represented as a finite order
differential operator acting on the RMT distribution function $P_{WD}(K)$
that corresponds to
$\Phi \left(
\alpha ,\beta \right) =1$ in Eq.~(\ref{s}): 
\begin{equation}
\label{differ}
P\left( k\right) =\Phi \left( -2\frac \partial {\partial a},-i\frac
\partial {\partial k}\right)\,\left.P_{a}(k)\right|_{a=1},
\end{equation}
where
\begin{eqnarray}
\label{Pa}
P_{a}(k)&=&a^{-1}P_{WD}(k/a)=\\ \nonumber
&=&\lim_{\phi \rightarrow
0}\frac{-\phi ^2}{32\pi \bar{g}}%
\Re\int DQ\exp \left[ -\frac \alpha 2a+ik\beta \right]  Str\widehat{k}%
Q_{0}^{11}Str\widehat{k}Q_{0}^{22}. 
\end{eqnarray}
Using the identities (see Eq.~\ref{ident} in Appendix \ref{app1}) 
that relate the derivatives of
$P_{a}(k)$ with
respect to $k$ to those with respect to $a$, we can rewrite $\delta
P(k)$ in the following form: 
\begin{equation}
\label{an}
\delta P\left( k\right) =
\Pi _4\left\{ \left( \frac 4d-1\right) \frac{\partial ^2}{%
\partial a^2}+\left( \frac 92-\frac{16}d+\frac{36}{d\left(
d+2\right) }%
\right) \frac \partial {\partial a}\right\}\left.
P_a\left( k\right) \right|_{a=1}.
\end{equation}
Note that by the definition given in Eq.(\ref{Pa}), the function $P_{a}(k)$
obeys two normalization conditions:
\begin{equation}
\label{nc}
\int dk\, P_{a}(k) =1;\;\;\;\int dk\,|k|\,P_{a}(k)=a.
\end{equation}
Using the first of these conditions we immediately conclude that
the cancellation of the terms proportional to $P_{a}(k)$ in Eq.(\ref{an})
ensures the conservation of the normalization $\int dk\,\delta P(K)=0$.

Next we note that the terms proportional to the first derivative can be
absorbed into the function $P_{a}(k)$: $P_{a=1}(k)+\delta
(\partial/\partial a)\,P_{a=1}(k)\approx P_{a=1+\delta}(k)$.
In doing so we observe with the help of the second normalization condition in
Eq.(\ref{nc}) that
$\langle|k|\rangle=\langle|K|\rangle/2\bar{g}=1+\delta$. Thus the terms
with the first derivative in Eq.(\ref{an}) result in a shift in the
average $\langle|K|\rangle$:
\begin{equation}
\delta\left\langle \left| K\right| \right\rangle =2\bar{g}\left(
\frac 92-\frac{16}d+\frac{36}{d\left( d+2\right) }\right) \Pi _4.
\end{equation}
By redefining the $k=K/\left\langle \left| K\right| \right\rangle $,
where $\left\langle \left| K\right| \right\rangle $ is the average of
absolute value of the level curvature, one can cancel the terms with the first
derivative in Eq.(\ref{an}). All what is left is the term with the second
derivative which describes the change in the {\it shape} of the
distribution function. The final result of these tedious calculations
is very simple:
\begin{equation}
\delta P(k)=\left(\frac{4}{d}-1
\right)\,\Pi_{4}\left.\frac{\partial^2}{\partial
a^2}\left(a^{-1}P_{WD}(k/a)\right)\right|_{a=1},\,\,\,\,\Pi_{4}\propto
1/\bar{g}^{2}.
\end{equation}
This equation means that the RMT curvature distribution $P_{WD}(k)$ plays
the role of the generating function for its own corrections.
\section{The signature of the pre-localized states in the 
level curvature distribution.}
It has been known for quite a while that the relaxation of current and the
local density of states (DOS)
in disordered conductors exhibit an anomaly even in the 
weak-localization regime. 
Namely, it has been shown in Ref.~[\onlinecite{b11}] that
there exists a small (but not exponentially small) probability 
of finding a current relaxation time or a local DOS 
that is much larger than the 
corresponding mean values. 
These anomalies have been attributed to quasi-localized
(or pre-localized) states \cite{b12,b13,b14}, that is, states with 
an anomalously large
peak in  $|\Psi({\bf r})|^2$ at some point ${\bf r}={\bf r}_{0}$.

Very recently the problem of current relaxation in disordered conductors has
been reconsidered \cite{b14,b15} by an elegant instanton
approximation \cite{b14} applied to the supersymmetric version of the nonlinear
sigma-model\cite{Efetov}. In these papers the main result of the previous
work \cite{b11} has been confirmed for $2d$ systems. However, the new
method was able to describe some unknown regimes of current relaxation and
to set correct limits of validity for the regimes found earlier. Later the
same idea \cite{b14} has been applied \cite{b12} to find directly the
distribution of $|\Psi ({\bf r})|^2$ and the distribution of local densities
of states \cite{b16}. Thus the existence of quasi-localized states has
been proved and the corresponding configuration of the random impurity
potential has been found\cite{b14,b17}.

The main idea of Edwards and Thouless \cite{Thouless74} is that it is
possible to distinguish
between localized and extended states by analyzing the sensitivity of the
spectrum to a twist of the phase in the boundary conditions. This sensitivity
is significant only for states with a localization radius larger than 
the sample size $L$
and negligible for strongly localized states. It is
clear that the existence of the pre-localized states should lead to an
enhancement in $P(K)$ at small $K$. In low dimensional systems $d=1,2$ 
where the pre-localized states correspond to localized states 
with an anomalously small localization radius,
one may expect a singularity in $P(K)$ at $K=0$. In $3d$ metal
the typical pre-localized state looks like a sharp peak in $|\Psi({\bf r})|^2$
on top of the extended background $|\Psi({\bf r})|^2 \sim L^{-d}=const$.
The level curvature that corresponds to such a state does not vanish but only
slightly decreases. Thus the pre-localized states in $3d$ 
in the weak localization regime
should have much weaker effect on the level curvature distribution.

\subsection{Instanton approximation.}

In order to check these predictions we consider, 
instead of $P(K)$ at small $K$,
its Fourier-transform
$\tilde{P}(\lambda)=\int dK\, P(K)\, e^{-iK\lambda}$ at $\lambda\gg 1$.

It is easy to see that for both the RMT result Eq.(\ref{RMT}) and 
the regular correction
Eq.(\ref{MR}) the function $\tilde{P}(\lambda)$ vanishes exponentially  
for $\lambda\gg 1$.
In what follows we will seek for slowly decreasing contributions. 
Support for the existence of
such contributions can be gained by noticing that in the functional 
representation of $P(K)$,
Eq.(\ref{main}), Eq.(\ref{F}),
the level curvature
$k$ plays the same role as the frequency in 
the problem of the current relaxation in disordered conductors
\cite{b14}, thus $\lambda$ being analogous to time. 
Therefore, one may expect long
nonexponential tails in $\tilde{P}(\lambda)$ in analogy with 
those present in the current 
relaxation function $I(t)$. However, it is far from clear 
that two problems are
equivalent, since the boundary conditions are different and 
the nonlinear sigma-model in Eq.(\ref{F}) contains
additional terms that describe the T-breaking perturbation.

The main idea of Ref.~[\onlinecite{b14}], that we will exploit here, is that at
large $\lambda$
the configurations of the field $Q({\bf r})$ that are space independent 
or slowly varying in space, are energetically unfavorable.
In contrast, essentially 
space-dependent configurations in the vicinity of the classical 
(instanton) solution
$Q_{ins}({\bf r})$ that minimizes the action $F[Q]$ appear
to be energetically advantageous.
At large $g$ the fluctuations around 
this solution are expected to be small and one arrives at:
\begin{eqnarray}
\tilde{P}(\lambda )\equiv Ae^{-S(\lambda )}=
\lim_{\phi \rightarrow 0}\Re \int {\cal D%
}Q\int_{-\infty }^{+\infty }dK\,A[Q_{ins};
\phi ]\,e^{-\{F[Q_{ins}]+iK\lambda \}}.
\label{Flam}
\end{eqnarray}
where $A[Q_{ins};\phi ]$ is a pre-exponential factor including the effect of 
fluctuations around the instanton solution.

The Grassmann variables in the action, Eq.(\ref{F}) can lead only to a
renormalization of the pre-exponential factor $A$ in Eq.(\ref{Flam}), since
the integration over these variables is equivalent to a differentiation.
Thus
with exponential accuracy we can neglect all the Grassmann variables in the
Efetov's parametrization for $Q({\bf r})$. Next, a finite contribution to $%
S(\lambda )$ in the limit $\phi \rightarrow 0$ comes only from the
infinitely large boson-boson components of the field $Q({\bf r})$. Therefore
we consider only the leading terms in the noncompact angles $\theta _1$ and 
$\theta _2$ in the Efetov's parametrization for the orthogonal ensemble: 
\begin{equation}
Q=V^{-1}HV,  \label{EP}
\end{equation}
where 
\begin{equation}
H=\left( \matrix{\cosh\theta_{B}& -\sinh\theta_{B}\cr \sinh\theta_{B}&
-\cosh\theta_{B}\cr}\right)_{R-A}\otimes P_{B},  \label{H}
\end{equation}
with $P_{B}=(\hat{k}+1)/2$,
\begin{eqnarray}
\cosh \theta _B &=&\cosh \theta _1\cosh \theta _2+\sigma _x\sinh \theta
_1\sinh \theta _2,  \label{Theta} \\
\sinh \theta _B &=&\sinh \theta _1\cosh \theta _2+\sigma _x\cosh \theta
_1\sinh \theta _2,  \nonumber
\end{eqnarray}
and 
\begin{equation}
V=e^{i\varphi \sigma _z}\otimes P_{+}\otimes P_B+e^{i\chi \sigma _z}\otimes
P_{-}\otimes P_B.  \label{V}
\end{equation}
The field $Q_{ins}({\bf r})$ must obey
periodic boundary conditions, since the twist of phase $\phi$ is taken 
into account explicitly in the action $F[Q]$. This means that 
$\theta_{1,2}({\bf r})$ and the functions 
$\exp[\varphi({\bf r})],\exp[\chi({\bf r})]$ 
should obey periodic boundary conditions.

In this way we obtain $F[Q_{ins}]=L^{-d}\int f[Q_{ins}]\,d^d {\bf r}$, where: 
\begin{eqnarray}  \label{FF}
& &f[Q]=\frac{\pi}{4}g\left[(\partial \theta_{+})^2+(\partial
\theta_{-})^2 \right]+ \\
&+&\frac{\pi}{2}g\left[(\partial\varphi-\phi{\bf n})^2+(\partial \chi)^2
\right](\cosh\theta_{+}\cosh\theta_{-}-1)-  \nonumber \\
&-&2\partial\chi\,(\partial\varphi-\phi{\bf n})\,\sinh\theta_{+}\sinh%
\theta_{-}-\frac{i\pi }{4}K\phi^2 (\cosh\theta_{+}+\cosh\theta_{-}). 
\nonumber
\end{eqnarray}
Here $\theta_{\pm}=(\theta_{1}\pm\theta_{2})/2$ with $\theta_{1,2}\geq 0$ and 
${\bf \partial}\equiv \{\partial/\partial 
x_{\alpha}\}$.

We will look for a minimum of the functional $F[Q_{ins}]+iK\lambda$ that
corresponds to: 
\begin{equation}  \label{cond}
\theta_{-}=\partial\chi=0,\;\;\;\; \theta_{+}\equiv\theta\in [0,+\infty].
\end{equation}
By varying the functional $F[Q_{ins}]+iK\lambda$ over $\theta$, $\varphi$
and $k$ we find: 
\begin{equation}  \label{1}
\partial^2\theta+\phi^2 [\kappa-(\partial v -{\bf n})^2] \sinh\theta=0,
\end{equation}
\begin{equation}  \label{2}
\partial\,[(\partial v-{\bf n}) (\cosh\theta-1)]=0,
\end{equation}
and 
\begin{equation}  \label{sc}
\frac{\pi}{4}\phi^2\int(\cosh\theta+1)\,d^d \rho=\lambda,
\end{equation}
where $\kappa=iK/2g$, $d^d\rho=\frac{d^d {\bf r}}{L^d}$ and $\varphi=\phi v$.
\\Eqs.(\ref{1})-(\ref{sc}) correspond to the global {\bf case I}. 
As usual, in the local 
{\bf case II} the terms linear in ${\bf n}=\{1,0,0,... \}$ are absent.

The limit $\phi\rightarrow 0$ is done simply by absorbing $\phi^2$ into $%
\theta$. We introduce $\tilde{\theta}=\theta+\ln\phi^2$. Then in the limit $%
\phi\rightarrow 0$ we have $\sinh\theta\approx\cosh\theta=\frac{1}{2}e^{\tilde{%
\theta}}\,\phi^{-2}$ and Eqs.(\ref{1}),(\ref{2}),(\ref{sc}) take the form: 
\begin{equation}  \label{11}
\partial^2\tilde{\theta}+\frac{1}{2}[\kappa-(\partial v -{\bf n})^2]\,e^{%
\tilde{\theta}}=0,
\end{equation}
\begin{equation}  \label{22}
\partial\,[(\partial v -{\bf n})\,e^{\tilde{\theta}}]=0,
\end{equation}
\begin{equation}  \label{ssc}
\frac{\pi}{8}\int e^{\tilde{\theta}}\, d^d \rho=\lambda.
\end{equation}
where now $\tilde\theta\in [-\infty,+\infty]$.

Using Eqs.(\ref{FF}),(\ref{cond}),(\ref{11}) and the periodicity of the
function $\tilde{\theta}({\bf r})$ we find: 
\begin{equation}  \label{S-l}
S(\lambda)=\frac{\pi}{4}g\int (\partial\tilde{\theta})^2\,d^d \rho +
2g\kappa\lambda.
\end{equation}

We note that $\phi$ drops from the problem only if we assume a {
\it topologically trivial} solution corresponding to periodic boundary 
conditions being imposed on $v({\bf r})$. 
Otherwise $\phi$ appears in the boundary 
condition for $v=\varphi({\bf r})/\phi$ which is periodic modulus $2\pi/\phi$
and thus is ill defined in the limit $\phi\rightarrow 0$. In what follows we consider 
only such a topologically trivial solution.

One can solve Eq.(\ref{22}): 
\begin{equation}  \label{sol}
(\partial v -{\bf n})=[\nabla\times{\bf A}]\,e^{-\tilde{\theta}},
\end{equation}
where $[\nabla\times{\bf A}]=const$ in $1d$ and is the $curl$ of an arbitrary 
vector function ${\bf A({\bf r})}$ in higher dimensions.
Below we consider only the simplest solution that corresponds to
$[\nabla\times{\bf A}]\equiv-{\bf n}/N=const$.

Let us consider first the local {\bf case II}. Doing the space 
integration of Eq.(\ref{sol})
which in this case does not contain the term proportional to ${\bf n}$,
and using periodic boundary conditions for $v({\bf r})$ one 
immediately arrives at $[\nabla\times {\bf A}]=\partial v=0$. Then the same
procedure with Eq.(\ref{11}) leads to the conclusion that the only
solution
for
$\tilde{\theta}$ that obeys periodic boundary conditions, is
space-independent and
exists only for $\kappa={\bf n}^2=1$. The
corresponding action is
$S(\lambda)=2g\lambda$. Thus the instanton approximation in the local 
{\bf case II} gives only an exponentially small tail
$\tilde{P}(\lambda)\propto
e^{-2g\lambda}$ that has been already obtained by the perturbative approach.  
We conclude that for {\bf case II}
the analogy with the problem of current relaxation
appears to be wrong.

Now consider the global {\bf case I}.
Integrating Eq.(\ref{sol}) over space and using the periodicity of $%
v({\bf r})$ gives: 
\begin{equation}  \label{N}
N=\int e^{-\tilde{\theta}}\,d^d \rho.
\end{equation}
Substituting Eqs.(\ref{sol}),(\ref{N}) into Eq.(\ref{11}) we finally arrive
at: 
\begin{equation}  \label{111}
\partial^2\tilde{\theta}+\frac{\partial
U}{\partial\tilde{\theta}}=\partial^2\tilde{\theta}
+\frac{\kappa}{2}\,e^{\tilde{\theta}}-
\frac{1}{2N^2}%
\,e^{-\tilde{\theta}}=0.
\end{equation}
It appears that the global nature of perturbation and the corresponding 
linear in ${\bf n}$ term in Eq.(\ref{sol}) leads to
a term proportional to $e^{-\tilde{\theta}}$ in Eq.(\ref{111}) that
builds a second ``wall'' in the effective ``potential'' $U(\tilde{\theta})$ 
and makes it possible for periodic solutions ("oscillations") to
exist.

Eq.(\ref{111}) takes a more symmetric form if we make a shift $\tilde{\theta}%
=u-\zeta$, where: 
\begin{equation}  \label{zeta}
\cosh\zeta=\left(\kappa+\frac{1}{N^2} \right)\,\frac{N}{2\sqrt{\kappa}}%
,\;\;\; \sinh\zeta=\left(\kappa-\frac{1}{N^2} \right)\,\frac{N}{2\sqrt{\kappa%
}}.
\end{equation}
Finally we have the system of equations: 
\begin{equation}  \label{a}
\partial^2 u+\gamma^2 \sinh u=0,
\end{equation}
\begin{equation}  \label{NN}
\frac{1}{N}=\gamma^2 \int e^{-u}\,d^d \rho,
\end{equation}
\begin{equation}  \label{l}
\lambda=\frac{\pi}{8\gamma^2 N^2}\,\int e^{u}\, d^d \rho,
\end{equation}
where $\gamma^2=\sqrt{\kappa}/N$.

Solving these equations for a
hyper-cubic sample $-1/2<\rho _i<1/2$ 
with periodic boundary conditions
one finds $u({\bf r},\lambda )$, $%
N(\lambda )$ and $\gamma (\lambda )$ which enter the instanton action $%
S(\lambda )$: 
\begin{equation}
S(\lambda )=\frac \pi 4g\int (\partial u)^2\,d^d\rho +2g\gamma ^4N^2\lambda .
\label{act}
\end{equation}

\subsection{Non-exponential tails of $\tilde{P}(\lambda)$ in 
low-dimensional systems.}

We will see below that for large $\lambda $ the parameter $\gamma $ is
small. For $\gamma \ll 1$ the term $\gamma ^2\sinh u$ is very small unless
$%
\sinh u$ is exponentially large. This means that we can approximate $\gamma
^2\sinh u\approx \frac{\gamma ^2}2e^{|u|}\,sign(u)$. 
Thus Eq.(\ref{a}) is replaced by the Liouville equation : 
\begin{equation}
\partial ^2u+\frac{\gamma ^2}2e^{|u|}\,sign(u)=0.  \label{Lio}
\end{equation}
The generic solution to the Liouville equation in the low-dimensional case $d=1,2$ 
is given in terms of two arbitrary 
functions $f(w)$ and $v(w)$ of the complex variable $w=i\gamma z/\sqrt{8}$
with $z=x+iy$:
\begin{equation}
e^{|u|}=\frac{2f^{^{\prime }}(w)\,v^{^{\prime }}(-w^{*})}{(f(w)+v(-w^{*}))^2},
\label{gen}
\end{equation}
where $f^{^{\prime }}=df/dw$ and $v^{^{\prime }}=dv/dw$. We need the r.h.s.
of Eq.(\ref{gen}) to be real positive. This can be done by the choice: 
\begin{equation}
f^{*}(w)v(-w^{*})=1.  \label{re}
\end{equation}
Then we have the solution in terms of one function only $F(z)=f(i\gamma z)$: 
\begin{equation}
e^{|u|}=\frac{16}{\gamma ^2}\frac{|\frac{dF}{dz}|^2}{(1+|F|^2)^2}.
\label{u-F}
\end{equation}

\subsubsection{Quasi-$1d$ Case}

Choosing  $F(z)=e^{kz+b}$ in Eq.(\ref{u-F}), one has 
a quasi-$1d$ solution to the Liouville equation 
that depends only on one coordinate
$x$:
\begin{equation}  \label{L1}
e^{|u|}=\frac{4k^2}{\gamma^2\,\cosh^2 (kx+b)},
\end{equation}
where $k$ and $b$ are real constants.

The solution on a ring $-\frac{1}{2}<x<\frac{1}{2}$
is constructed by reflecting anti-symmetrically 
the positive solution with $b=0$ 
around the points $x=\pm\frac{1}{4}$ 
[see  Fig.~\ref{fig2}]. 
The second constant $k$ is found from the condition $u(\pm 1/4)=0$:
\begin{equation}  \label{eq-k}
4k^2=\gamma^2\,\cosh^2 (k/4), \;\;\;\; k\approx \ln(1/\gamma^4).
\end{equation}
The anti-symmetry of the solution immediately leads to the identity:
\begin{equation}
\label{sym}
I=\int e^{u}\,dx=\int e^{-u}\,dx.
\end{equation}
Since the function $u\sim \ln(1/\gamma^2)$ is large everywhere except
in the vicinity of its zeros at $|x|=\frac{1}{4}$ we have:
\begin{equation}  \label{I}
I\approx \int_{-\frac{1}{4}}^{+\frac{1}{4}} e^{u}\,dx=\frac{8k}{%
\gamma^2}\,\tanh(k/4)\approx \frac{8k}{\gamma^2}.
\end{equation}
Next we calculate the integral:
\begin{equation}  \label{1part}
\int_{-\frac{1}{2}}^{+\frac{1}{2}}(u^{^{\prime}})^2 \,dx=2\int_{-\frac{1}{4}%
}^{+\frac{1}{4}}(u^{^{\prime}})^2 \,dx=4k^2-16 k\approx 4k^2.
\end{equation}
Then from Eqs.(\ref{NN}),(\ref{l}) we have:
\begin{equation}  \label{1N}
N=\frac{1}{8k},
\end{equation}
\begin{equation}  \label{1sc}
\gamma^4 = \frac{64\pi k^3}{\lambda}.
\end{equation}
Finally using Eq.(\ref{act}),(\ref{eq-k}) and (\ref{1sc}) we arrive at:
\begin{equation}  \label{1S}
S(\lambda)=\pi g k^2 \approx \pi g \ln^2 \lambda.
\end{equation}
Thus the characteristic function in a quasi-$1d$ systems is:
\begin{equation}  \label{11F}
\tilde{P}(\lambda)=
A \exp\left[-\frac{g_{1}}{2}\ln^2 \lambda\right],\;\;\;\;g_{1}=2\pi g,
\end{equation}
The above result holds 
within the domain of validity of the nonlinear sigma-model Eq.(\ref{F}). 
This model 
and hence the saddle-point  
equations work only for sufficiently slow varying fields $Q({\bf r})$,
namely
 $|\partial u|<L/l$, where $l$ is the elastic
scattering length. It follows immediately from Eqs.(\ref{L1}),(\ref{eq-k})
that the above result is valid for $1\ll \lambda\ll exp(L/l)$.

The logarithmically-normal tail in $\tilde{P}(\lambda)$ 
described by Eq.(\ref{11F})
is exactly of the same functional form as the current 
relaxation function $I(t)$
in Ref.~[\onlinecite{b14}] for the orthogonal ensemble.

\subsubsection{$2d$ Case}

In full analogy with the quasi-$1d$ case, we construct  
a double-periodic solution to the Liouville equation
on a torus $-\frac{1}{2}<x,y<\frac{1}{2}$ by reflection.
We consider a positive solution $u(z)$ inside the square
$\Omega$ with vertices at  
$z=\pm 1/2,\pm i/2$ and 
then continue it anti-symmetrically about a side of the
square in any quarter of the sample 
$|\Re z|<1/2$, $|\Im z|<1/2$. The definition domain of the solution 
with its sign is drawn in Fig.~\ref{fig3}.
By construction, the symmetry relationship
Eq.(\ref{sym}) is valid for such a $2d$ solution too. 

The procedure of finding the solution is described in the Appendix \ref{app3}.
We note that for our purposes we need only the solution for $|z|=r\ll 1$.
It is rotationally invariant and has the form:
\begin{equation}  \label{rad} 
e^{u(r)}=\frac{16b(k-1)^2\,r^{2k-4}}{(\gamma^2+b\,r^{2k-2})^2},
\end{equation}
where
\begin{equation}  \label{b}
b=16\left( \frac{\pi^2}{2}\right)^{k}(k-1)^2.
\end{equation}
Note that the solution Eq.(\ref{rad}) can be immediately obtained from the
radial Liouville equation, with $k$ and $b$ being two constants of
integration. The requirement of periodicity of $u(z)$ helps to establish
a connection, Eq.(\ref{b}), between these constants. 

The remaining
constant $k$
is found in a standard way from the requirement of convergence of $\int
(\partial u)^2 dxdy$ in the action $S(\lambda)$: 
\begin{equation}  \label{gr u}
(\partial u)^2= \left[\frac{2k-4}{r}-4b\,(k-1)\frac{r^{2k-3}}{%
\gamma^2+br^{2k-2}}\right]^2.
\end{equation}
Thus we immediately find: 
\begin{equation}  \label{k}
k=2,\;\;\;\; b=4\pi^4,
\end{equation}
and 
\begin{equation}  \label{nabla}
(\partial u)^2=\frac{16 b^2 r^2}{(\gamma^2+b r^2)^2}.
\end{equation}
Because of the symmetry of $u(z)$, the integral $\int_{\Omega}(\partial
u)^2\,d^2\rho$ over the square $\Omega$ is exactly one-half of the total
integral over the period (over the sample) $\int (\partial u)^2\,d^2\rho$.
It diverges logarithmically at $r\gg \gamma$, and we arrive at: 
\begin{equation}  \label{nabint}
\int (\partial u)^2\,d^2\rho=32\pi \ln\left(\frac{Cb}{\gamma^2} \right),
\end{equation}
where $C=\frac{2}{\pi^2 e}$ can be found from an exact solution in the
region $|z|\sim 1$. 

The result is almost independent of $b$
at small $\gamma$ and is essentially determined by the logarithmic
solution of the Poisson equation that follows from Eq.(\ref{a}) at
$\gamma=0$.

Now let us calculate the integrals in the self-consistency equations. For
symmetry reasons we have: 
\begin{equation}  \label{sci}
\int e^{u}\,dxdy=\int e^{-u}\,dxdy=\frac{16\pi\,(k-1)}{\gamma^2}=\frac{16\pi%
}{\gamma^2}.
\end{equation}
Most of the contribution to these integrals comes form the 
small $r\sim\gamma\ll 1$ 
and the result is
independent of $b$.

Now we are in the position to calculate the constants $\gamma$ and $N$ that
enter the instanton action, Eq.(\ref{act}). They are given by Eqs.(\ref{NN}%
),(\ref{l}),(\ref{sci}): 
\begin{equation}  \label{Na}
N=\frac{1}{16\pi},\;\;\;\; \gamma^4=\frac{8^3\pi^4}{\lambda}.
\end{equation}
Then the final expression for the instanton action in $2d$ reads: 
\begin{equation}  \label{ia}
S(\lambda)=4\pi^2 g \left[\ln\left(\frac{\lambda}{8}\right)-1 \right].
\end{equation}
Accordingly, the characteristic function $\tilde{P}(\lambda)$ turns out
to have a power-law asymptotic behavior at large $\lambda\gg 1$: 
\begin{equation}  \label{PL}
\tilde{P}(\lambda)=A\left(\frac{c}{\lambda}
\right)^{2g_{2}},\;\;\;\;g_{2}=2\pi^2 g,
\end{equation}
where $c=8e$.

Few notes should be made on the validity of the result Eq.(\ref{PL}).
Firstly, the above instanton approximation with the action $S(\lambda)$
logarithmic in $\lambda$
is only justified when $g\gg 1$, since the pre-exponential factor $A$
could also be a power-law function of $\lambda$ but with an exponent of
order 1. Secondly, the nonlinear sigma-model and hence the saddle-point
equations work only for $|\partial u|<L/l$, where $l$ is the elastic
scattering length. It follows immediately from Eq.(\ref{nabla}),(\ref{Na})
that the above result is valid for $1\ll \lambda\ll (L/l)^4$.

\subsection{Nonanalyticity of the level curvature distribution.}

In this section we show that the slowly decreasing tails in the
characteristic
function
$\tilde{P}(\lambda)$ at $\lambda\gg 1$ given by Eqs.(\ref{11F}),
(\ref{PL}) result in a
nonanalytic behavior of $P(K)$ at $K=0$. As usual, true
nonanalyticity
arises only in the 
thermodynamic limit $L/l\rightarrow\infty$ 
\cite{note4},
since only in this limit the tails extend
to infinity. For any finite $L/l$ the function $P(K)$ is still 
analytic at $K=0$ but the region of the regular behavior
of $P(K)$ shrinks to zero with increasing $L/l$. Below
the limit $L/l\rightarrow\infty$ is assumed.

Let us consider the {\bf quasi-$1d$} case first. In this case all
derivatives of $P(K)$ are finite at $K=0$:
\begin{equation}
P^{(2n)}(0)=\int_{-\infty}^{+\infty}(-1)^{n}\lambda^{2n}\,\tilde{P}(\lambda)
\frac{d\lambda}{2\pi}
\propto
\exp [(2n+1)^2/2g_1]. \label{der}
\end{equation}
Yet the function $P(K)$ is {\it nonanalytical} at K=0, since the
Taylor series $P(K)=\sum_{n}\frac{P^{(2n)}(0)}{(2n)!}K^{2n}$ has zero
radius of convergence because of the very fast growth of $P^{(2n)}(0)$
with $n$.
 
The singularity at $K=0$ is much stronger in {\bf $2d$ case}.
In this case all derivatives $P^{(2n)}(0)$ with $2n+1>2g_{2}$ 
are proportional to $(L/l)^{4(2n+1-2g_{2})}$ and diverge in the
thermodynamic limit. Let us define $m$ as an integer
obeying the inequality of $|g_{2}-m|\leq\frac{1}{2}$. Then the
the expansion of $P(K)$ at small $K$ has the form:
\begin{equation}
\label{expans}
P(K)=c_{0}+c_{1}K^2 + ...
c_{m-1}K^{2(m-1)}+c_{m}K^{2m-\alpha_{m}}+o(K^{2m}),
\end{equation}
where $0<\alpha_{m}<2$ is given by:
\begin{equation}
\alpha_{n}=(2n+1)-2g_{2}.
\end{equation}

In the {\bf $3d$} metal case we failed to find a solution to the
saddle-point
problem that would lead to a finite action $S(\lambda)$ in the
thermodynamic limit. This means that the characteristic function
$\tilde{P}(\lambda)$ has only regular corrections at $g\gg 1$
and thus decays exponentially for $\lambda\gg 1$.

\section{Non-analyticity of $P(K)$ at $K=0$ and multifractality of
eigenfunctions.}

From the results of the previous section
we see that the strength of the singularity of $P(K)$
at $K=0$ depends on the dimensionality in a nonmonotonic way.
The singularity is very weak in a quasi-$1d$ metal; it 
reaches a maximum in a $2d$ metal
where $P(K)$ has a branch cut;
it disappears in a $3d$ metal where
the level curvature distribution is analytic.

Such a behavior is related to the fact that $d=2$ is the low critical
dimension for the Anderson transition, and the wavefunctions in the $2d$
weak-localization regime share some features of the critical wavefunctions
at the Anderson transition in higher dimensions.

\subsection{$P(K)$ at the Anderson transition in $2+\epsilon$ dimensions.}

The usual way to describe the critical state near the Anderson transition
is the $(d-2)=\epsilon$-expansion. To this end one considers the
quantity of interest in a $2d$ system with $g_{2}\gg 1$ and
then replaces $g_{2}$ by the critical conductance $g_{d}^{*}=1/(d-2)$
which is
the fixed point
of the scaling equation \cite{a21,note5}:
\begin{equation}  \label{SE}
\frac{d\ln g_{d}}{d\ln L}=(d-2)-\frac{1}{g_{d}}+o\left(\frac{1}{g_{d}^2}
\right).
\end{equation}
For the orthogonal ensemble in $d=2+\epsilon$ dimensions we find to
the leading order in
$\epsilon\ll 1$:
\begin{equation}
\label{crit}
\tilde{P}_{c}(\lambda)\propto
\left(\frac{1}{\lambda}\right)^{\frac{2}{\epsilon}}.
\end{equation}
Note that at the critical point, the conductance $g_{d}^{*}$ is {\it exactly}
size-independent, and one can consider the thermodynamic limit
$L\rightarrow\infty$ without tuning other parameters in order to keep
$g_{d}^{*}$ fixed.

Thus one can define the critical exponent $\mu$ that determines the
power-law tail of the critical characteristic function:
\begin{equation}
\label{mu}
\tilde{P}_{c}(\lambda)\propto\lambda^{-\mu},\qquad
\mu=\frac{2}{\epsilon}+o(1).
\end{equation}
If we set $\epsilon=1$ in the above equations we find $\mu\approx 2$.
Then it follows from Eq.(\ref{expans}) that
already the second derivative of $P(K)$ at $K=0$ is divergent:
\begin{equation}
\label{cr3}
P_{c}(K)=c_{0}-c_{1}|K|^{2-\alpha}, \qquad \alpha=3-\mu.
\end{equation}

\subsection{Exponent $\mu$ and multifractality.}

Unfortunately it is known that the accuracy of the $d-2=\epsilon$
expansion is quite poor and insufficient for a precise determination 
of the critical exponents. In this situation one can try to find relationships
between different critical exponents rather than try to evaluate them
using the $\epsilon$-expansion. This certainly requires some {\it
assumptions} about the underlying physics.

As has been mentioned in the Introduction, a unique property of the
critical states is multifractality. This property is characterized by the
{\it power-law} dependences of averaged powers of eigenfunction amplitudes
$|\Psi_{E}({\bf r})|$. Two of such
power-law dependences are known \cite{b8,b9}. One of them
determines the scaling of a {\it single} eigenfunction with
respect to the size of the system $L$.
\begin{equation}
\label{IPR}
\sum_{{\bf r},n}\langle|\Psi_{n}({\bf
r})|^{2q}\delta(E-E_{n})\rangle\;\propto
L^{-d_{q}(q-1)},
\end{equation} 
Another one determines the correlations of {\it different}
eigenfunctions as a function of energy difference:
\begin{equation}
\label{IPC}
\sum_{{\bf r},n,m}\langle|\Psi_{n}({\bf
r})|^{q}\;|\Psi_{m}({\bf
r})|^{q}\delta(E-E_{n})\,\delta(E'-E_{m})\rangle\;\propto
|E-E'|^{-(1-\frac{d_{q}}{d})\,(q-1)}.
\end{equation}
In Eqs.(\ref{IPR}),(\ref{IPC})  $d_{q}<d$ is a fractal
dimension that depends on $q$ (``{\it multifractality}'').

It is remarkable that Eq.(\ref{IPR}) can be derived \cite{b12} for the case
of $2d$ metals
by means of an instanton approximation 
similar to the one we used in this paper. The spectrum of the 
fractal dimensions $d_{q}$ obtained in this approximation turns out to be
linear\cite{note6}:
\begin{equation}
\label{sfd}
d_{q}=d-\frac{\eta}{2}\,q, \qquad \eta=d-d_{2}=\frac{2}{\beta g_{2}},
\end{equation}
where $\beta=1,2,4$ for the orthogonal, unitary and symplectic ensembles.

It is reasonable to assume that the power-law tail in $\tilde{P}(\lambda)$
is another signature of multifractality \cite{CanBasu}. Then one may hope
that the expression for $\mu$ in terms of the structural constant of
multifractality $\eta=d-d_{2}$ provides a better approximation for
$\mu$ than the $\epsilon$-expansion. By using Eq.(\ref{sfd}) and the
relationship\cite{note7}
$\mu=2\beta g_{2}$, we obtain:
\begin{equation}
\label{sr}
\mu=\frac{4}{\eta}.
\end{equation}
The derivation of Eq.(\ref{sr}) that we have just carried out for
the $d=2$ case is based on two crucial facts: 
i). the exponent $\mu$ is determined by the spectrum of multifractality $d_{q}$ 
and ii). this spectrum is linear (for $q\ll 1/\eta$).
We will now make the assumption that i). is valid for any critical state.
Since for any critical state with weak
multifractality the spectrum of $d_{q}$ is expected to be
linear up to very large values of $q$, we believe that Eq.(\ref{sr}) is
valid for
{\it any}
critical
state with weak multifractality. In contrast to Eq.(\ref{mu}), the
relationship between $\mu$ and $\eta$ Eq.(\ref{sr}) is independent of
dimensionality and
the symmetry parameter $\beta$ and should apply to $2d$ critical states
in the Quantum Hall regime and for systems with spin-orbit interaction 
\cite{b8,b9}.

\section{Numerical evaluation of $P(K)$ for the Anderson model.}
For our numerical analysis we consider a tight-binding model on a square
lattice of $L^{d}$ sites. The one-particle Hamiltonian is:
\begin{equation}
\label{tbm}
H=\sum_{i}\epsilon_{i}c^{\dagger}_{i}c_{i}+t\sum_{\langle<ij\rangle>}
(e^{i\theta_{ij}}\,c^{\dagger}_{i}c_{j}+
e^{-i\theta_{ij}}\,c^{\dagger}_{j}c_{i}).
\end{equation}
The site energies $\epsilon_{i}$ are randomly distributed with uniform
probability between $-W/2$ and $W/2$ in units of $t=1$. 
The parameter $W$ controls the
amount of disorder in the system. The phase shifts $\theta_{ij}$ in the
hopping term connecting nearest neighbors represent the effect of
an external perturbation that breaks the T-invariance of the system.
As for the analytical calculations we consider two types of such
perturbations. The first one ({\bf case I}) is the usual Aharonov-Bohm
flux $\Phi=(\phi/2\pi)\Phi_{0}$ 
that pierces the system closed to a ring geometry giving rise
to a $\it global$ shift of the boundary conditions in one direction.
We will choose a gauge such that each hop in $x$-direction picks up a
phase $\theta_{ij}=\phi/L$, so that total twist of the boundary condition
is $\phi$. The second one ({\bf case II}) is a random magnetic flux.
In this case the gauge is such that the phase $\theta_{ij}$ relative to a
hop in the $x$-direction is Gaussian distributed with zero average and
variance equal to $\langle\theta_{ij}^2 \rangle=(\phi/L)^2$. For this
gauge, the vector-potential ${\bf A}(i)\propto\{\theta_{i,i+1},0,...\}$ is
defined on
the dual lattice with sites in the middle of bonds in $x$-direction.
Thus this is a random vector-potential model with a short-range correlator
$\langle A_{\alpha}(i)A_{\beta}(j)\rangle=(\Phi/L)^2 \,
\delta_{\alpha,x}\delta_{\beta,x}\,\delta_{ij}$ 
of the type given in the continuous approximation by Eq.(\ref{vp}).
The only difference is that the correlator is ${\it anisotropic}$.  
This difference is not important, since it leads only to a constant
factor $1/d$ in $v_{\tau}$ that can be absorbed in the parameter $\phi$.
This kind of perturbation is qualitatively different from {\bf case I},
since it acts {\it locally}.

The numerical evaluation of the curvature is based on the representation
of the second derivative by the finite difference:
\begin{equation}
\label{fd}
K_{n}=2\phi^{-2}\,[E_{n}(\phi)-E_{n}(0)].
\end{equation}
In using this formula one should take care that $\phi$ is small enough
in order for Eq.(\ref{fd}) to be valid. On the other hand, a $\phi$ too small
would result in big numerical errors because of the finite numerical precision
in evaluating $E_{n}$. 
The optimal choice of $\phi$ should be made
for each level $E_n$ separately, since the level curvatures vary in a wide
range for a given realization of disorder. To this end, we diagonalize the
Hamiltonian for several values of $\phi$ (up to ten values)
for each disorder realization and choose
the smallest $\phi$ for which the normalized difference
$|E_{n}(\phi)-E_{n}(0)|/E_{n}(0)$ is still larger than some conveniently
chosen small parameter.
In order to attain a smooth curve and decrease the statistical
fluctuations the
statistical average has been made both over the energies (in the energy window
of width $4$ centered at $E=0$) and over many realizations of the
disorder (there were typically few thousands of them). 

\subsection{Corrections to $P(K)$ beyond RMT in $3d$ metals.}

In this section we compute the finite $g$ corrections to the shape of the
curvature distribution in the metallic regime, comparing the numerical
results with the regular corrections, Eq.(\ref{MR}),(\ref{C}).

It turns out that the magnitude of corrections is small and we need to
consider a rather large disorder ($W\sim 10$) to detect it. 
At yet larger values of disorder we may enter the critical regime.
The onset of the critical regime
exhibits itself in the weak dependence of the conductance $g$
and the magnitude of correction to $P(k)$ on the system size $L$. 
In contrast in a good metal
a naive estimation of $C_{d}$ in Eq.(\ref{MR}) yields
$|C_{d}|\propto W^4 /L^2$. The upper limit of the coefficient $|C_{d}|$
can be estimated from the critical conductance $g_{3}^{*}=4\pi^2
g^{*}\approx 1$. Since $|C_{d}|$ is proportional to a small parameter
$(4\pi^3 g)^{-2}=(\pi g_{3})^{-2}$, its value just near the critical
region is proportional to a small numerical factor $1/\pi^2 \sim 0.1$.
It is this small numerical factor together with the strong $W$-dependence
of $C_{d}$ that makes the correction to $P(k)$ small and Eq.(\ref{MR})
qualitatively applicable even very close to the mobility edge.
 
\subsubsection{Global Vector Potential.}

In Fig.~\ref{fig4} we show the numerical results for
$\delta P(k)=P(k)-P_{WD}(k)$
for the $3d$ metallic regime in {\bf case I}. The calculations are performed
for system size $L=8$ and disorder $W=12$. The number of disorder
realizations is 1500. The deviation of $P(k)$ from the RMT result is very
small, less than one percent. The magnitude of the statistical noise
present after averaging is done appears to be only a little smaller than
the
signal itself. Nevertheless the general trend of the curve agrees with 
the analytical prediction Eq.(\ref{MR}): $P(k)$ is {\it above} the RMT
result at small $k$. We have used the coefficient $C_{d}$
in Eq.(\ref{MR})
as a free parameter in the least square fitting of the numerical results.
The value $C_{3}=0.0044$ found from such a fitting is probably a reliable
estimate of the magnitude of the correction $\delta P(k)$ in the above
case.

\subsubsection{Random Magnetic Flux}

The same correction $\delta P(k)$ for the case of a random magnetic flux
is displayed in Fig.~\ref{fig5}. 
The values of the parameters of the Hamiltonian
are the same as for the previous case. Despite the statistical
fluctuations are still rather strong, the numerical results are quite
significant. We see that again the expression Eq.(\ref{MR}) provides a
rather good one-parameter fitting function for the numerical results.
However, in this case the coefficient $C_{d}=-0.014$ is {\it negative}
in full agreement with the analytical prediction. Moreover, 
the numerical results are consistent with the analytical prediction even
{\it quantitatively}.
It follows from Eq.(\ref{C}) that there is a magic relationship for the
ratio of amplitudes of the correction in {\bf case I} and {\bf case II}:
\begin{equation}
\label{magrat}
R=\frac{C_{3}^{(I)}}{C_{3}^{(II)}}=-\frac{1}{3},
\end{equation}
Our calculations give a result $R=-0.32$ which is in an amazingly good
agreement with Eq.(\ref{magrat}).

\subsection{$P(k)$ at the mobility edge in $3d$ and in $2d$ metals.}

A numerical investigation of the distribution $P(k)$ at the Anderson
transition critical point has been already carried out in
Ref.~[\onlinecite{CanBasu}]. The main finding of the numerical simulation 
is that the distribution function at the mobility edge is remarkably well
fitted by the formula:
\begin{equation}
\label{pcrit}
P_{\alpha}(k)=\frac{A_{\alpha}}{(1+|k|^{(2-\alpha)})^{\frac{3}{2-\alpha}}}.
\end{equation}
with $\alpha\approx 0.4$. Equation (\ref{pcrit}) defines a function that has
a branching point of the type Eq.(\ref{cr3}) at $k=0$ and the asymptotic 
behavior $P_{\alpha}(k)\propto |k|^{-3}$, expected in all cases
where there is energy level repulsion $R(\omega,0)\propto |\omega|$
at $\omega\ll 1$.  The function $P_{\alpha}(k)$ is a rather
special one \cite{CanBasu}, since once it is properly normalized by
choosing
$A_{\alpha}= (2-\alpha)\Gamma[3/(2-\alpha)]
\{[\Gamma[1/(2-\alpha)]\Gamma(2/(2-\alpha)]\}^{-1}>1$, it 
automatically satisfies the condition $\int dk |k| P_{\alpha}(k)=1$ 
for {\it all} $\alpha$. Thus the full distribution $P_{\alpha}(k)$
is determined uniquely by its value at $k=0$. If we take $P_{\alpha}(k=0)$
equal to the numerical result (that is known up to a small error bar)
the whole curve is parameter free.
Alternatively we can consider $\alpha$ as a free parameter that should 
be determined by a least square fitting of the overall numerical curve.
These two procedures yield very close values for $\alpha$ and an excellent
overall fitting of the numerical curve, implying that the extrapolation
by means of the function
$P_{\alpha}(k)$ is very self-consistent.

In Fig.~\ref{fig6} we plot the results for the difference
$\delta P(k)=P(k)-P_{WD}(k)$ for
the critical disorder $W=16.5$ and the system size $L=12$ as compared to
two one-parameter fitting curves provided by Eq.(\ref{MR}) and
Eq.(\ref{pcrit}). It is clearly seen that despite the analytic function
$\delta P_{\rm reg}(k)$ given by Eq.(\ref{MR}) reproduces a correct
qualitative behavior of
$\delta P(k)$,
there is some feature at small $|k|$ that is captured better by the
nonanalytic fitting function, Eq.(\ref{pcrit}). 

We note also that this enhancement at small $|k|$ relative to $\delta
P_{\rm reg}(k)$ is size-dependent, with $P(k=0)$ increasing with $L$ so that a
full saturation is not reached even for $L=10$. 
Such a behavior is beyond the one-parameter scaling and can be explained
by the termination of the power-law
tail in the
characteristic function                      
$\tilde{P}(\lambda)$ at a finite $\lambda =
(L/l)^4$. 
Indeed, the power-law tail given by Eq.(\ref{mu}),
makes a contribution to $P(k=0)$ proportional to
$\int_{1}^{(L/l)^4}\,\lambda^{-\mu}\,
d\lambda$. This contribution is size-dependent and increases with
increasing $L$ (with a saturation at $L/l\rightarrow\infty$) even exactly
at the
critical point where the exponent $\mu$ is a constant.

Now let us calculate the fractal dimension $d_{2}=3-\eta$ using
the relationship, Eq.(\ref{sr}), between $\eta$ and $\mu=3-\alpha$.
For $\alpha\approx 0.4$ we have $\mu\approx 2.6$ and $d_{2}\approx 1.5$.
This value is in a good agreement with direct evaluation \cite{b24} of
$d_{2}$
from Eqs.(\ref{IPR}),(\ref{IPC}).

We also checked that in a $2d$ metal the deviation from the RMT result
$\delta P(k)$ has the same qualitative form as in the $3d$ critical case.
In Fig.~\ref{fig7} we present plots of $\delta P(k)$ 
for different values of $L$ and $W=6$. An interesting feature seen
in the figure is
the fixed point in $\delta P(k)$ at $k \approx 0.35$.
Despite the one-parameter function $P_{\alpha}(k)$ does not have
an {\it exact} fixed point at the same value of $k$, all the curves
obtained varying $\alpha$ do get very close to each other at $k \approx 0.35$
in what looks almost like a fixed point.
As in the $3d$ critical case,
the analytic function $\delta P_{\rm reg}(k)$ fits well the overall
distribution but fails to describe the sharp enhancement for small
curvatures that is instead well described by the function
$P_{\alpha}(k)$. This is shown in Fig.~\ref{fig8} for the system
$L=30, W=6$.
 
\section{Conclusions}
In this paper we have investigated both analytically and numerically
the relationship between the
statistics of eigenfunctions and the spectral
statistics in disordered conductors. The level curvature distribution 
has been chosen as the target
of our study, since it is the simplest known example of parametric
spectral statistics that can be used as a spectral probe
of the structure of eigenfunctions.
\\The main results of the paper are formulated in Eqs.(\ref{MR}),(\ref{C}),
and Eqs.(\ref{cr3}),(\ref{sr}). Numerical results in agreement with
these analytical predictions are presented
in Figs.~\ref{fig4},\ref{fig5}, and \ref{fig6},\ref{fig8} respectively.
The first two equations describe the regular corrections beyond RMT to the
level curvature distribution in disordered metals. 
These corrections stem from
long-range correlations in the wave functions with a typical length scale
of the order of the sample size.
The latter two equations summarize the effect of local
irregularities (sharp peaks) in the structure of eigenfunctions
in its most developed form (multifractality) in the critical region near
the mobility edge. 

Moreover, Eq.(\ref{sr}) suggests an explicit relationship
between the fractal dimension $d_{2}=d-\eta$ of a critical eigenfunction 
and the exponent $\mu$ 
in the power-law tail Eq.(\ref{mu}) of the characteristic function
$\tilde{P}_{c}(\lambda)$
that describes the particular parametric spectral statistics 
considered here (level curvature distribution).

Note that Eq.(\ref{sr}) is more general than Eq.(\ref{cr3}). The latter
requires rather strong multifractality $\eta>\frac{4}{3}$, while the
former applies to a generic critical state. For instance, it would be
interesting to check its validity for the critical state in the quantum
Hall effect, where $\eta\approx 0.5$ and we predict $\mu\approx 8$.
Recent progress \cite{Koeln} in numerical simulations on the
Chalker-Coddington network model \cite{ChalCod} seems to make the task
attainable.  

As far as the regular corrections are concerned, there is an interesting
question of what happens to them for $d>4$. The sum in Eq.(\ref{C}) is
a parameter-free number only for $d<4$
when it converges. For $d>4$ the sum is divergent and requires a cut-off
at large $|{\bf q}|$. Thus for the correct evaluation of this sum 
it is necessary to go beyond the diffusion approximation and the
approximation of slow spatial variations of the field $Q({\bf r})$
in the nonlinear sigma-model. The divergent sum in
Eq.(\ref{C}) implies that for $d>4$ the correction $\delta P(k)$ to the level
curvature distribution is dominated by {\it short-range}
spatial correlations of the eigenfunctions, in contrast to
$d<4$ where it is dominated by {\it long-range}
correlations. Based on our discussion at the end of Sec.~\ref{mres},
the short-range nature of the eigenfunction
correlations in $d>4$ is most likely the cause of the change of sign
in the correction $\delta P(k)$ for the
"global" case I (as compared to $d<4$) and the reason why $\delta P(k)$
shows qualitatively the same behavior
of case II
("local" perturbations).

One of the most important results Eq.(\ref{r}) of our calculations is that
the ratio $r(g)$ of the mean level curvature $\langle|K|\rangle$ and the mean
Drude conductance $2g$ is not a constant and is always larger than the RMT
result $r=1$.
This is in a qualitative agreement with the result \cite{A-L,Akk}
that  $\langle|K|\rangle\propto \sqrt{g}$ in a strictly one-dimensional
case where $g\ll 1$ and localization effects
are strong. Indeed, let
us assume that the square-root dependence is typical for strongly
localized states in any dimensions.
Then the function $r(g)$ should behave like
$r(g)\propto g^{-1/2}$ at small $g\ll 1$ and $r(g)\rightarrow 1$ for 
$g\rightarrow\infty$.
If in addition we make the natural assumption that $r(g)$ is a monotonic
function, we arrive at the conclusion that $\delta r(g)>0$ everywhere
in agreement with Eq.(\ref{r}).
 
A similar deviation from the proportionality relationship
$\langle|K|\rangle\propto g$
has been observed recently in numerical simulations \cite{BHMM}.

{\bf Acknowledgements}\\ 
We thank B.L.Altshuler, E.Akkermans,  
V.I.Fal'ko, Y.V.Fyodorov, I.V.Lerner, A.D.Mirlin, W.~Stephan and Yu Lu for 
stimulating discussions.
V.E.K. is grateful for the hospitality extended to him 
at the Newton Institute (Cambridge, UK) where the final part of this work
has been completed.
Support from grants RFBR/INTAS No.95-675,
CRDF No.RP1-209 (V.E.K.)
and EPSRC grant No.GR/K95505 (I.V.Y.)
is also gratefully acknowledged.
C.M.C. thanks the Swedish NFR and TFR for financial support.

\appendix
\section{Effective Action}
\label{app1}

The parametrization (\ref{separ}) enables us to single out 
fast modes $\widetilde{%
Q}$ in the action (\ref{F}) as follows: 
\begin{equation}
F\left[ \widetilde{Q},Q_0\right] =Str\left\{ \frac{\pi g}8\left( \left[ 
\overrightarrow{Q}_\phi ,\widetilde{Q}\right] -iL\nabla \widetilde{Q}\right)
^2-\left( i\frac{\pi g}4k\phi ^2Q_\Lambda +Q_J\right) \widetilde{Q}\right\} ,
\end{equation}
where the following notation has been used: 
\[
\overrightarrow{Q}_\phi =T_0\widehat{\phi }T_0^{-1},\qquad Q_\Lambda
=T_0\Lambda T_0^{-1},\qquad Q_J=T_0JT_0^{-1}. 
\]
\[
g=\frac D{\Delta L^2}=\nu DL^{d-2},\quad STr...\equiv \int \frac{d{\bf r}}%
VStr...\ ,\quad K=2g\cdot k. 
\]
Taking advantage of the fact that we are interested in $\phi \rightarrow
0 $ limit only, it is convenient to absorb the large conductance 
into the definition of the flux $\frac
\pi 2g\phi ^2\rightarrow \phi ^2$ (wherever possible). Then introducing
the dimensionless gradient $\partial =\sqrt{\pi /2}L\nabla $ and 
using the identity 
\[
\frac{\phi ^2}2STrQ_\Lambda \widetilde{Q}=STr\overrightarrow{Q}_\phi ^2%
\widetilde{Q}, 
\]
we recast the action in the following form 
\begin{equation}
F\left[ \widetilde{Q},Q_0\right] =STr\left\{ \frac 14\left( \left[ 
\overrightarrow{Q}_\phi ,\widetilde{Q}\right] -i\sqrt{g}\partial \widetilde{Q%
}\right) ^2-\left( ik\overrightarrow{Q}_\phi ^2+Q_J\right) \widetilde{Q}%
\right\} .  \label{F1}
\end{equation}
These manipulations, along with fact that $\widetilde{W}$ scales as $%
\widetilde{W}=g^{-1/2}w$, $w\propto 1$ enable us to construct 
a perturbative expansion in the parameter $1/g-$%
straightforwardly, simply expanding $F\left[ \widetilde{Q},Q_0\right] $ in
powers of $w$. This procedure is equivalent to selecting diagrams contributing
to the same order in $1/g$. Then for the partition function we get the
representation: 
\begin{equation}
Z=\lim_{\phi \rightarrow 0}\frac{-\phi ^2}{32\pi g }\Re\int
DQ_0D\widetilde{Q}%
\cdot J\left[ \widetilde{Q}\right] \exp \left( -F\left[ \widetilde{Q}%
,Q_0\right] \right) ,  \label{Z}
\end{equation}
where $J\left[ \widetilde{Q}\right] $ is the Jacobian of the transformation $%
Q\rightarrow \left( \widetilde{Q},Q_0\right) $(obtained in the 
following Appendix \ref{app2}).
The action (\ref{F1}) is expanded as follows: 
\begin{equation}
F\left[ \widetilde{Q},Q_0\right] =F\left[ Q_0\right] +\frac 14STr\left(
\partial w\right) ^2+\sum_{n=1}^\infty g^{-n/2}F_{n/2}\left[ Q_0,w\right] .
\label{exp}
\end{equation}
The first term of the expansion is nothing but the zero-mode action 
($RMT$-limit): 
\[
F\left[ Q_0\right] =STr\left[ \frac 12\left( \overrightarrow{Q}_\phi \Lambda
\right) ^2-\left( ik\overrightarrow{Q}_\phi ^2+Q_J\right) \Lambda \right] . 
\]
The second term in (\ref{exp}) corresponds to the noninteracting diffusion
modes approximation. The other terms in (\ref{exp}) describe the interaction of
the diffusion and zero modes with $g^{-1}$ playing the role of a coupling
constant. For our
purposes it is enough to keep the first four interaction terms of expansion: 
\begin{equation}
F_{1/2}=iSTr\overrightarrow{Q}_\phi w\partial w,  \label{v1/2}
\end{equation}
\begin{eqnarray}
F_1 &=&\frac 12STr\left\{ A\Lambda w^2+\left( \overrightarrow{Q}_\phi
\Lambda
w\right) ^2+\frac 14\left( \partial w\right) ^2w^2\right\}+\\
\nonumber&+&\frac 1{8i}%
STr\left\{\overrightarrow{Q}_\phi \left( \partial w^3-2w\cdot \partial
w\cdot w\right)
\right\} ,  \label{v1}
\end{eqnarray}
\begin{eqnarray}
F_{3/2}&=&\frac 14STr\left\{ A\Lambda w^3+2\left( \overrightarrow{Q}_\phi
\Lambda w\right) ^2w\right\}+\\ \nonumber
&+&\frac{1}{4i}STr\left\{\overrightarrow{Q}_\phi \left(
\partial w^3\cdot
w+\partial w\cdot w^3+\partial w^2\cdot w^2\right) \right\} ,  \label{v3/2}
\end{eqnarray}
\begin{eqnarray}
&& F_2 =\frac 18STr\left\{ A\Lambda w^4+\left( \overrightarrow{Q}_\phi
\Lambda w^2\right) ^2+2\left( \overrightarrow{Q}_\phi \Lambda w\right)
^2w^2+\frac 1{2i}\overrightarrow{Q}_\phi \partial w^5\right\} ,  \label{v2}
\\
&&\ \ \ +\frac 1{64}STr\left\{ \left( \partial w^3\right) ^2+2\partial
w^5\cdot \partial w-2\partial w^4\cdot \partial w^2\right\} .  \nonumber
\end{eqnarray}
We have used the notation: 
\begin{equation}
A=\overrightarrow{Q}_\phi \Lambda \overrightarrow{Q}_\phi -ik\overrightarrow{%
Q}_\phi ^2-Q_J.
\end{equation}
Averaging over the fast modes can be implemented by the use of contraction rules
derived in \cite{AKL86}. We recast them in the following form: 
\begin{eqnarray}
&&\left\langle \widetilde{W}\left( {\bf r}\right) R\widetilde{W}\left(
{\bf r}%
^{\prime }\right) \right\rangle =\Pi \left( {\bf r},{\bf r}^{\prime
}\right) \left\{ STrR-STr\Lambda R\Lambda -\overline{R}+\Lambda
\overline{R}%
\Lambda \right\} ,  \label{cr} \\ &&
\left\langle \widetilde{W}\left( {\bf r}\right) STrR\widetilde{W}\left( {\bf %
r}^{\prime }\right) \right\rangle =\Pi \left( {\bf r},{\bf r}^{\prime
}\right) \left\{ R-\overline{R}-\Lambda \left( R-\overline{R}\right) \Lambda
\right\} .  \nonumber
\end{eqnarray}
The propagator $\Pi \left( {\bf r},{\bf r}^{\prime }\right) $ satisfies
the diffusion equation: 
\begin{equation}
-\Delta \Pi \left( {\bf r},{\bf r}^{\prime }\right) =\frac 1{\pi gL^2}\cdot
\delta \left( {\bf r}-{\bf r}^{\prime }\right) ,\quad \Pi ^{-1}\left( {\bf q}%
\right) =\pi g\left( {\bf q}L\right) ^2.
\end{equation}
The contraction rules (\ref{cr}) provide a basis for integrating out 
the fast modes
in (\ref{Z}). For example, straightforward but lengthy calculations give us
the following rules for the integration of product of 
two vertices from (\ref{v1}) containing no gradients:
\begin{eqnarray*}
&&\left\langle STrA\Lambda \widetilde{W}^2\left( {\bf r}\right)
STrA\Lambda 
\widetilde{W}^2\left( {\bf r}^{\prime }\right) \right\rangle _c =4\Pi
^2\left( {\bf r},{\bf r}^{\prime }\right) \left\{ \left[ STrA\Lambda \right]
^2-\left[ STrA\right] ^2\right\} , \\
&&\left\langle STr\left[ B\Lambda \widetilde{W}\left( {\bf r}\right)
\right]
^2STr\left[ B\Lambda \widetilde{W}\left( {\bf r}^{\prime }\right) \right]
^2\right\rangle _c =4\Pi ^2\left( {\bf r},{\bf r}^{\prime }\right) \cdot
\left\{ \left[ STr\left( B\Lambda \right) ^2\right] ^2\right.+\\ 
&+& \left.\left[
STrB^2\right]
^2 -2\left[ STrB^2\Lambda \right] ^2+STr\left( \left(
B\Lambda \right)
^4+B^4-2\left( B^2\Lambda \right) ^2\right) \right\},\\
&& 
\left\langle STr\left[ B\Lambda \widetilde{W}\left( {\bf r}\right) \right]
^2STrA\Lambda \widetilde{W}^2\left( {\bf r}^{\prime }\right) \right\rangle
_c=\\ &=& 4\Pi ^2\left( {\bf r},{\bf r}^{\prime }\right) \cdot STrA\left(
2B^2\Lambda -B\Lambda B-\Lambda \left( B\Lambda \right) ^2\right) . 
\end{eqnarray*}
where the matrices satisfy the symmetry 
relations $\overline{A}=A,\quad 
\overline{B}=-B$ (for notations see \cite{Efetov})and the brackets $\left\langle
_{...}\right\rangle _c$ mean keeping only the connected parts of the correlators
after averaging over fast modes $\widetilde{W}$ .

These calculations can be easily extended to couplings of gradient vertices
as well. Then having at hand all possible couplings we are able to construct
a cumulant expansion for the partition function (\ref{Z}). Keeping all
nonvanishing correlators up to the second order in $1/g$ we arrive at: 
\begin{eqnarray}
F^{eff} &=&F_0+\left\langle F_1\right\rangle -\frac 12\left\langle
F_1^2\right\rangle _c+\frac 12\left\langle F_1F_{1/2}^2\right\rangle
_c-\frac 1{4!}\left\langle F_{1/2}^4\right\rangle _c, \\
F_1^{eff} &=&\left\langle F_1\right\rangle =-\frac 12\sum\limits_{{\bf q}%
}\Pi \left( {\bf q}\right) \cdot STr\left( \overrightarrow{Q}_\phi \Lambda
\right) ^2,  \nonumber
\end{eqnarray}
\[
F_2^{eff}=-\frac 12\left\langle F_1^2\right\rangle _c+\frac 12\left\langle
F_1F_{1/2}^2\right\rangle _c-\frac 1{4!}\left\langle F_{1/2}^4\right\rangle
_c=4\sum\limits_{{\bf q}}\Pi ^2\left( {\bf q}\right) \cdot \left(
f_2+f_3+f_4\right) . 
\]
Here the term $f_2$ corresponds to coupling of two gradientless vertices (%
\ref{v1}): 
\begin{eqnarray*}
&&-8f_2 =\left[ STrA\Lambda \right] ^2-\left[ STrA\right] ^2+\left[
STr\left( \overrightarrow{Q}_\phi \Lambda \right) ^2\right] ^2-2\left[ STr%
\overrightarrow{Q}_\phi ^2\Lambda \right] ^2 \\
&&-STr\left\{ \left( \overrightarrow{Q}_\phi \Lambda \right)
^4-2\left( 
\overrightarrow{Q}_\phi ^2\Lambda \right) ^2+2A\left( 2\overrightarrow{Q}%
_\phi ^2\Lambda -\overrightarrow{Q}_\phi \Lambda \overrightarrow{Q}_\phi
-\Lambda \left( \overrightarrow{Q}_\phi \Lambda \right) ^2\right) \right\} ,
\end{eqnarray*}
$f_3$ describes the coupling of two gradient vertices (\ref{v1/2}) to the
gradientless one(\ref{v1}): 
\begin{eqnarray*}
2d\cdot f_3 &=&STrA\Lambda \cdot STr\left( \overrightarrow{Q}_\phi \Lambda
\right) ^2-2STrA\cdot STr\overrightarrow{Q}_\phi ^2\Lambda +\left[ STr\left( 
\overrightarrow{Q}_\phi \Lambda \right) ^2\right] ^2 \\
&&\ \ -4\left[ STr\overrightarrow{Q}_\phi ^2\Lambda \right] ^2+2STr\left( 
\overrightarrow{Q}_\phi ^2\Lambda \right) ^2,
\end{eqnarray*}
and the last contribution $f_4$ comes from 4 coupled gradient vertices (\ref
{v1/2}): 
\[
-\frac 49d\left( d+2\right) \cdot f_4=\left[ STr\left( \overrightarrow{Q}%
_\phi \Lambda \right) ^2\right] ^2-4\left[ STr\overrightarrow{Q}_\phi
^2\Lambda \right] ^2 
\]

The expression for $F^{eff}$ can be simplified considerably if one uses
the following factorization properties of long traces:
\begin{eqnarray}
\label{fac}
&&Str[(\hat{\phi}^2 Q_{0})^2]=\left[Str[\hat{\phi}^2
Q_{0}]\right]^2=\beta^2\\ 
\nonumber
&&Str[(\hat{\phi} Q_{0})^4]=\frac{1}{2}\left[Str[(\hat{\phi}
Q_{0})^2]\right]^2=\frac{\alpha^2}{2}\\
\nonumber
&&Str[(\hat{\phi} Q_{0})^2
\hat{\phi^2}Q_{0}]=\frac{1}{2}Str[(\hat{\phi}
Q_{0})^2]\,Str[\hat{\phi}^2
Q_{0}]=\frac{\alpha\beta}{2}\\
\nonumber
&&Str[(\hat{\phi} Q_{0})^2
\hat{k}\,Q_{0}^{pp}]=\frac{1}{2}Str[(\hat{\phi}
Q_{0})^2]\,Str[\hat{k}\, Q_{0}^{pp}]=\frac{\alpha}{2}\,Str[\hat{k}\,
Q_{0}^{pp}].
\end{eqnarray}
The
factorization holds in the leading power in the noncompact angles
in the original Efetov's parametrization and can be shown by
straightforward but extremely lengthy calculations.

It is worth noting that the Jacobian $J\left[ \widetilde{Q}\right] $
also contributes to the effective action (the last term in Eq.(\ref{f2})).
Taking into account Eq.(\ref{ja})
of Appendix \ref{app2} one concludes that the Jacobian leads to the
replacement ${\bf q}^2 +\frac{1}{\pi g}$ in $\Pi({\bf q})$. Thus
the $1/g^2$ contribution from the Jacobian follows from the correction to
$\Pi({\bf q})$ in $F_{1}^{eff}$:  
\begin{equation}
F_{jacob}^{eff}=\frac 12\sum_{{\bf q}}\Pi \left( {\bf q}\right) \cdot
STr\left( \overrightarrow{Q}_\phi \Lambda \right) ^2  \label{jacob}
\end{equation}
Collecting all the results obtained in this Appendix we arrive at the
effective action ${\cal F}\left[ Q_0\right] $ 
given in Eqs.(\ref{f2}),(\ref{fj}).

Using these equations one can represent $P(k)$ in the differential form
of Eq.(\ref{differ}). It turns out that all the relevant derivatives of
$P_{a}(k)$ over $k$ can
be expressed  through the derivatives over $a$. Using the
following identities:
\begin{eqnarray}
&&\left.k\frac{\partial}{\partial k}P_{a}(k)\right|_{a=1}=
-\left.\left(1+\frac{\partial}{\partial
a}\right)\,P_{a}(k)\right|_{a=1};\\ \nonumber
&&\left.k\frac{\partial^2}{\partial k\partial a}P_{a}(k)\right|_{a=1}=
-\left.\left( 
\frac{\partial^2}{\partial  
a^2}+2\frac{\partial}{\partial a}\right)\,P_{a}(k)\right|_{a=1};\\
\nonumber
&&\left.\frac{\partial^2}{\partial k^2}P_{a}(k)\right|_{a=1}=\left.
\left(\frac{\partial}{\partial a}-\frac{\partial^2}{\partial
a^2}\right)P_{a}(k)\right|_{a=1};\\ \nonumber
&&\left.k^2\frac{\partial^2}{\partial k^2}P_{a}(k)\right|_{a=1}=\left.
\left(2+4\frac{\partial}{\partial a}+\frac{\partial^2}{\partial
a^2}\right)P_{a}(k)\right|_{a=1}
\label{ident}
\end{eqnarray}
we can finally represent $\delta P(k)$
in the form Eq.(\ref{an}) that contains only the first and second
derivatives with respect of $a$.

\section{Calculation of the Jacobian}
\label{app2}

In the evaluation of the Jacobian of the transformation 
$Q\rightarrow \left( Q_0,\widetilde{Q}\right) $
we follow the procedure proposed in 
Ref.~[\onlinecite{FyodMir95}] and
prove that the Jacobian does not depend on zero-mode $Q_0$\cite{mistake}.
The derivation is simplified if we go to the ``rational'' parametrization
first: 
\begin{equation}
Q=\left( 1-W\right) \Lambda \left( 1-W\right) ^{-1},\quad W=\left( 
\begin{array}{cc}
& B \\ 
\overline{B} & 
\end{array}
\right) .  \label{rat}
\end{equation}
This parametrization has been known since Efetov's work \cite{Efetov}. The
advantage of this representation lies in the fact that Jacobian of the
transformation $Q\rightarrow W$ equals to unity. On the another hand we
have the decomposition (\ref{separ}). Taking into account that $T_0$ belongs to
the graded coset space $UOSP\left( 2,2|4\right) /UOSP\left( 2|2\right)
\otimes UOSP\left( 2|2\right) $ the complete parametrization takes the form: 
\begin{eqnarray}
Q &=&T_0^{-1}\widetilde{Q}T_0,\quad T_0=\sqrt{\frac{1+W_0}{1-W_0}},\quad 
\widetilde{Q}=\left( 1-\widetilde{W}\right) \Lambda \left( 1-\widetilde{W}%
\right) ^{-1},  \label{dec} \\
W_0 &=&\left( 
\begin{array}{cc}
& B_0 \\ 
\overline{B}_0 & 
\end{array}
\right) ,\quad \widetilde{W}=\left( 
\begin{array}{cc}
& \widetilde{B} \\ 
\overline{\widetilde{B}} & 
\end{array}
\right) .  \nonumber
\end{eqnarray}
Comparing these two representations (\ref{rat}) and (\ref{dec}) we derive
the connection between $W$ and $\left( W_0,\widetilde{W}\right) $: 
\begin{eqnarray}
W &=&\left( W_0+\varpi \right) \left( 1+W_0\varpi \right) ^{-1},\;\; 
\varpi
=\left( 1-W_0^2\right) ^{-1/2}\widetilde{W}\left( 1-W_0^2\right) ^{1/2},
\label{sub} \\
B &=&\left( B_0+b\right) \left( 1+B_0b\right) ^{-1},\quad b=\left( 1-B_0%
\overline{B}_0\right) ^{-1/2}\widetilde{B}\left( 1-\overline{B}_0B_0\right)
^{1/2}.  \nonumber
\end{eqnarray}
Since the field $b$ (being proportional to $\widetilde{W}$) 
contains only nonzero
momenta it can be treated perturbatively. Going to the Fourier
representation and expanding (\ref{sub}) up to second order in $b$ we
separate zero- and nonzero momenta: 
\begin{eqnarray}
B\left( {\bf k}=0\right) &=&B_0-Sb_{{\bf q}}\overline{B}_0b_{-{\bf q}}, \\
B\left( {\bf k}\neq 0\right) &=&S\left( b_{{\bf k}}-b_{{\bf k}+{\bf q}}%
\overline{B}_0b_{-{\bf q}}+b_{{\bf k}+{\bf q}_1+{\bf q}_2}\overline{B}_0b_{-%
{\bf q}_1}\overline{B}_0b_{-{\bf q}_2}\right) ,  \nonumber
\end{eqnarray}
where $S=\left( 1-B_0\overline{B}_0\right) $ has been introduced and
the summation over repeated indecis is implied. Therefore, we are interested in
Jacobian of the transformation $B\rightarrow \left( B_0,b\right) $ which is
equivalent to $\left( B\left( {\bf k}=0\right) ,B\left( {\bf k}\neq 0\right)
\right) \rightarrow \left( B_0,b_{{\bf k}\neq 0}\right) $. The corresponding
Jacobian can be represented in the block-matrix structure: 
\begin{equation}
J=SDet\left( 
\begin{array}{cc}
\partial B\left( 0\right) /\partial B_0 & \partial B\left( 0\right)
/\partial b \\ 
\partial B\left( \emptyset \right) /\partial B_0 & \partial B\left(
\emptyset \right) /\partial b
\end{array}
\right) ,  \label{eqJ}
\end{equation}
where a short notation for the following supermatrices has been introduced:
\begin{eqnarray*}
\left[ \partial B\left( 0\right) /\partial b\right] _{0k} &=&\partial
B\left( q=0\right) /\partial b_k, \\
\left[ \partial B\left( \emptyset \right) /\partial B_0\right] _{k0}
&=&\partial B\left( k\neq 0\right) /\partial B_0, \\
\left[ \partial B\left( \emptyset \right) /\partial b\right] _{kk^{\prime }}
&=&\partial B\left( k\neq 0\right) /\partial b_{k^{\prime }},
\end{eqnarray*}
and right derivatives are implied \cite{Efetov}. Using the identity for
the superdeterminant of block matrices we recast (\ref{eqJ}) in the
following form 
\begin{eqnarray}
J &=&J_1\cdot J_2,\qquad J_1=SDet\frac{\partial B\left( \emptyset \right) }{%
\partial b}, \\
J_2 &=&S\det \left[ \frac{\partial B\left( 0\right) }{\partial B_0}-\frac{%
\partial B\left( 0\right) }{\partial b_k}\left( \frac{\partial B\left(
\emptyset \right) }{\partial b}\right) _{kk^{\prime }}^{-1}\frac{\partial
B\left( k^{\prime }\neq 0\right) }{\partial B_0}\right] ,  \nonumber
\end{eqnarray}
where $S\det $ acts within the space of $4\times 4$ matrices, 
while $SDet$ spans $k$%
-space also. Lengthy but straightforward calculations give us following very
useful formulae:

\[
Str\frac{\partial \left( DbF\right) }{\partial b}=StrD\cdot StrF,\quad Str%
\frac{\partial \left( D\overline{b}F\right) }{\partial b}=StrD\overline{F}, 
\]
\[
Str\frac{\partial \left( D_1b_1F_1\right) }{\partial b_1}\frac{\partial
\left( D_2b_2F_2\right) }{\partial b_2}=StrD_1D_2\cdot StrF_1F_2, 
\]
\begin{equation}
Str\frac{\partial \left( D_1b_1F_1\right) }{\partial b_1}\frac{\partial
\left( D_2\overline{b}_2F_2\right) }{\partial b_2}=StrD_1D_2\overline{F}_1%
\overline{F}_2,  \label{usf}
\end{equation}
\[
Str\frac{\partial \left( D_1b_1F_1\right) }{\partial b_1}\left[ \frac{%
\partial \left( Gb\right) }{\partial b}\right] ^{-1}\frac{\partial \left( D_2%
\overline{b}_2F_2\right) }{\partial b_2}=StrD_1G^{-1}D_2\overline{F}_1%
\overline{F}_2. 
\]

Now we expand $J_1$ up to second order in $b$: 
\[
J_1=SDet\frac{\partial \left( Sb\right) }{\partial b}\cdot K,
\]
where
\[ 
K=\exp
\left\{
STr\left( \frac{\partial \left( Sb\right) }{\partial b}\right) ^{-1}\frac{%
\partial \left( b\overline{B}_0b\overline{B}_0b\right) }{\partial
b}-\frac
12STr\left[ \left( \frac{\partial \left( Sb\right) }{\partial b}\right) ^{-1}%
\frac{\partial \left( Sb\overline{B}_0b\right) }{\partial b}\right]
^2\right\}. 
\]
Taking into account that $SDet\left[ \partial \left( Sb\right) /\partial
b\right] =1$ and exploiting (\ref{usf}) we obtain: 
\begin{equation}
Str\left( \frac{\partial \left( Sb\right) }{\partial b}\right) ^{-1}\frac{%
\partial \left( Sb\overline{B}_0b\overline{B}_0b\right) }{\partial b}%
=\sum\limits_{{\bf q}\neq 0,{\bf k}}Strb_{{\bf q}}\overline{B}_0Str\overline{%
B}_0b_{-{\bf q}},
\end{equation}
\begin{equation}
Str\left[ \left( \frac{\partial \left( Sb\right) }{\partial b}\right) ^{-1}%
\frac{\partial \left( Sb\overline{B}_0b\right) }{\partial b}\right]
^2=2\sum\limits_{{\bf k}^{\prime }\neq 0,{\bf k}\neq 0}Strb_{{\bf k}^{\prime
}-{\bf k}}\overline{B}_0Str\overline{B}_0b_{{\bf k}-{\bf k}^{\prime }}.
\end{equation}
Finally, we arrive at: 
\[
J_1=\exp \left\{ \sum_{{\bf k}}Strb_{-{\bf k}}\overline{B}_0Str\overline{B}%
_0b_{{\bf k}}\right\} . 
\]
Then with the same accuracy we may recast $J_2$ in the following form: 
\begin{eqnarray*}
J_2 &=&S\det \left( 1-\delta J_2\right) , \\
\delta J_2 &=&\frac{\partial \left( Sb_{{\bf q}}\overline{B}_0b_{-{\bf q}%
}\right) }{\partial B_0}-\frac{\partial \left( Sb_{{\bf q}}\overline{B}_0b_{-%
{\bf q}}\right) }{\partial b_{{\bf k}}}\left( \frac{\partial \left(
Sb\right) }{\partial b}\right) ^{-1}\frac{\partial \left( Sb_{{\bf k}%
}\right) }{\partial B_0} \\
-\ln J_2 &\approx &Str\frac{\partial \left( Sb_{{\bf q}}\overline{B}_0b_{-%
{\bf q}}\right) }{\partial B_0}-Str\frac{\partial \left( Sb_{{\bf q}}%
\overline{B}_0b_{-{\bf q}}\right) }{\partial b_{{\bf k}}}\left( \frac{%
\partial \left( Sb\right) }{\partial b}\right) ^{-1}\frac{\partial \left(
Sb_{{\bf k}}\right) }{\partial B_0}
\end{eqnarray*}
Using formulae (\ref{usf}) we obtain: 
\[
Str\frac{\partial \left( Sb_{{\bf q}}\overline{B}_0b_{-{\bf q}}\right) }{%
\partial B_0}=StrSb_{{\bf q}}\overline{b}_{-{\bf q}}-Str\overline{B}_0b_{%
{\bf q}}\overline{B}_0b_{-{\bf q}} 
\]
\begin{eqnarray*}
&&\ \ Str\frac{\partial \left( Sb_{{\bf q}}\overline{B}_0b_{-{\bf q}}\right) 
}{\partial b_{{\bf k}}}\left( \frac{\partial \left( Sb\right) }{\partial b}%
\right) ^{-1}\frac{\partial \left( Sb_{{\bf k}}\right) }{\partial B_0} \\
\ &=&-StrSb_{-{\bf k}}\left( 1-\overline{B}_0B_0\right) ^{-1}\overline{b}_{%
{\bf k}}-Str\overline{B}_0b_{{\bf k}}Str\overline{B}_0b_{-{\bf
k}}\\  \ &+&StrSb_{-%
{\bf k}}\overline{b}_{{\bf k}}-Str\overline{B}_0b_{{\bf k}}\overline{B}_0b_{-%
{\bf k}}
\end{eqnarray*}
As a result we find for the second contribution $J_2$: 
\begin{eqnarray*}
-\ln J_2 &\approx &Str\left( 1-B_0\overline{B}_0\right) b_{-{\bf k}}\left( 1-%
\overline{B}_0B_0\right) ^{-1}\overline{b}_{{\bf k}}+Str\overline{B}_0b_{%
{\bf k}}Str\overline{B}_0b_{-{\bf k}} \\
\ &=&\frac 12\int \frac{dr}VStr\widetilde{W}^2+Str\overline{B}_0b_{{\bf k}%
}Str\overline{B}_0b_{-{\bf k}}
\end{eqnarray*}
The parametrization used in this Appendix differs slightly from the one
used in the main body of paper. To come back to the original parametrization we
must substitute $\widetilde{W}\rightarrow \widetilde{W}/2$ . Then collecting
contributions from $J_1$ and $J_2$ together we finally obtain Jacobian in
the form: 
\begin{equation}
J\left[ \widetilde{Q}\right] =\exp \left[ -\frac 18\int \frac{dr}VStr%
\widetilde{W}^2\right] .  \label{ja}
\end{equation}
As one can see from Eq.~(\ref{ja}) 
the Jacobian does not contain the zero-modes at all. Morover, being
quadratic, it just plays the role of a small frequency $\left( \sim 1/g\right)$
in the free diffusion propagator. Expanding such modified propagator we recover
the Jacobian contribution to the effective action Eq.(\ref{jacob}).

\section{Solution to  {\it 2d} Liouville equation}
\label{app3}

On the boundary of the square $\Omega$ [see Fig.~\ref{fig3}] the
function $u(z)$ and its first derivative must be continuous.
Because, by contruction, $u(x)$ is anti-symmetric when reflected
around the sides of $\Omega$, 
it must be zero on the boundary of $\Omega$:
\begin{equation}
\left| \frac{dF}{dz}\right| ^2=\frac{\gamma ^2}{16}(1+|F(z)|^2)^2,
\label{eq}
\end{equation}
For the case $\gamma \rightarrow 0$ we look for a solution $|F(z)|\gg 1$.
Then we have
\begin{equation}
\frac d{dz}\left( \frac 1{F(z)}\right) =\frac \gamma 4\,\exp(i\Theta
),\;\;\;\;Im\Theta =0.  \label{Th}
\end{equation}
Suppose we manage to find $\Theta (z)$ which is an analytic function
inside the square $\Omega $ and is real on its boundary. Then the equation
for $1/F$ is trivially reducible to quadratures and the solution $F(z)$
is an analytic
function inside the square $\Omega $. Then in the limit
$\gamma\ll 1$ the solution to the Liouville equation that obeys the
condition $u(z)=0$ on the boundary of the square $\Omega$ is found by
substituting the function $F(z)$ into Eq.(\ref{u-F}). 
In particular, in the region where $|F(z)|\gg 1$ we have:
\begin{equation}
e^{|u(z)|}=|e^{i\Theta (z)}|^2. \label{mod} \end{equation}
Consider the function
\begin{equation}
z(t)=\frac{\sqrt{2}}\pi \int_0^t\frac{d\xi }{(1-\xi ^4)^{1/4}}
\label{conf}
\end{equation}  
which does the conformal transformation of the unit circle in the complex
plane of $t$ onto the square $\Omega $ in the complex plane of $z$. This
function obeys the symmetry property:
\begin{equation}
z(it)=iz(t),\;\;\;\;z(t^{*})=z^{*}(t).  \label{Ssym}
\end{equation}  
If we choose $e^{i\Theta (z)}=[t(z)]^{-k}$, where $k$ is a real parameter,
the analytic function $\Theta (z)$ will automatically be real on the
boundary of the square $\Omega $. Integrating Eq.(\ref{Th}) we arrive at:
\begin{equation}
\frac 1{F(z)}=\frac \gamma 4\int_0^z[t(z^{\prime })]^{-k}\,dz^{\prime }.
\label{1/F}
\end{equation}
For $|z|\ll 1$ we have from Eq.(\ref{conf}):
\begin{equation}
z=\frac{\sqrt{2}}{\pi}\,t
\end{equation}
and
\begin{equation}  \label{F(z)}
F(z)= -\frac{4}{\gamma}\left(\frac{\pi}{\sqrt{2}}
\right)^{k}(k-1)\,z^{k-1},
\;\;\;\;\; (|z|\ll 1).
\end{equation}
We see that, indeed, $|F(z)|\gg 1$ for
\begin{equation}  \label{r0}
|z|>r_{0}=\frac{\sqrt{2}}{\pi}\,\left[\frac{\gamma}{2\pi\sqrt{2}(k-1)}
\right]^{\frac{1}{k-1}},\;\;\;\;(k>1).
\end{equation}
In this region the function $u(z)$ is given by:
\begin{equation}  \label{eu}
u(z)=-2k\ln |t(z)|=-2k\, \Re\ln t(z),\;\;\;\;(|z|\gg r_{0}).
\end{equation}
 
Thus for
$r_{0}\ll 1$, the solution to the $2d$ Liouville equation with 
boundary conditions $u(z)=0$
on the boundary of the square $\Omega$  
is given by Eq.(\ref
{u-F}) with $F(z)$ defined in Eq.(\ref{1/F}). For $|z|\ll 1$ this
solution for $u(z)$
depends only on $|z|=r$ and is given by Eq.(\ref{rad}).

\begin{figure}
\caption{Perturbative $1/g$ correction to the curvature distribution
$P(k)$ beyond the RMT result $P_{\rm WD}(k)$.
The expression for  $\delta P_{\rm reg}(k) = P(k) - P_{\rm WD}(k)$ is
given in Eqs.~(\ref{MR}),(\ref{C}). The dashed line represents the
case of global $T$-breaking perturbation (Case I). The solid line represents
the case of local $T$-breaking perturbation (Case II).}
\label{fig1}
\end{figure}

\begin{figure}
\caption{Plot of the periodic solution $u(x)$ of the Liouville equation
Eq.~(\ref{Lio}), 
that enters the instanton action of Eq.~(\ref{act}),
for the quasi-$1$D case. The function is defined on the ring 
$-\frac{1}{2} < x < \frac{1}{2}$. 
In the interval $-\frac{1}{4} < x < \frac{1}{4}$ the function $u(x)$ 
is the positive solution
of Eq.~(\ref{L1}) with the choice $b=0$ and 
$k$ found from the condition $u(\pm 1/4) =0$. 
In the intervals
$\pm\frac{1}{2} < x < \pm\frac{1}{4}$ \
$u(x)$ is constructed by antisymmetric
reflection around the points $\pm \frac{1}{4}$.}
\label{fig2}
\end{figure}

\begin{figure}
\caption{Definition domain of the periodic solution $u(z)$ of the Liouville
equation, Eq.~(\ref{Lio}), for the $2d$ case. The domain is a torus
$-1/2<x,y<1/2$, is represented here by the large square.
Inside the square $\Omega$ with vertices at $z= \pm1/2,\  \pm i/2$ the
function $u(z)$ is the real positive solution of  Eq.~(\ref{Lio}). 
In the remaining part of the larger square, $u(z)$ is constructed by
anti-symmetric reflection around the sides of the square $\Omega$
and its sign is negative.}
\label{fig3}
\end{figure}

\begin{figure}
\caption{Numerical results for $\delta P(k)=  P(k)-P_{\rm WD}(k)$
for the $3d$ Anderson model in the metallic regime.
The disorder is $w= 12t$ and the system size is L=12.
The curvatures 
are calculated for case I, global vector potential. The smooth
curve is fit with the analytical result given in Eqs.~(\ref{MR}),(\ref{C}).
The coefficient $C_3$ is taken as a free parameter
in the least square fitting.}
\label{fig4}
\end{figure}

\begin{figure}
\caption{The same as in Fig.~\ref{fig4} but for case II, local random 
magnetic field.}
\label{fig5}
\end{figure}

\begin{figure}
\caption{Numerical results for the difference
$\delta P(k)=P(k)-P_{WD}(k)$ for the $3d$ Anderson model of system size $L=12$
at the Anderson critical point (disorder $W=16.5$).
The dashed and solid curves are
two one-parameter fitting curves provided by
by Eq.(\ref{MR}) and Eq.(\ref{pcrit}).}
\label{fig6}
\end{figure}

\begin{figure}
\caption{Numerical results for the difference
$\delta P(k)=P(k)-P_{WD}(k)$ for the $2d$ Anderson model of different 
system sizes
$L=12,16,20,24,30$ and site disorder $W=6$.}
\label{fig7}
\end{figure}

\begin{figure}
\caption{Numerical results for the difference
$\delta P(k)=P(k)-P_{WD}(k)$ for the $2d$ Anderson model of system size $L=30$
and site disorder $W=6$. The dashed and solid curves are
two one-parameter fitting curves provided by
by Eq.(\ref{MR}) and Eq.(\ref{pcrit}).} 
\label{fig8}
\end{figure}


\begin{thebibliography}{999}
\bibitem{note1} {This statement 
is not true \cite{KravZirn} in the case where there is a degeneracy of
spectrum at $\phi=0$ (which is lifted for $\phi\neq 0$) for different
values of T-odd quantities such as spin
(Kramers degeneracy) or angular momentum (in perfect systems).}
\bibitem{note2}  {In this paper we have
changed sign in the definition of $STr$ compared to the original Efetov's 
paper\cite{Efetov} and Ref.~\onlinecite{b22}. 
We assume $STr A= Tr A_{BB}-Tr A_{FF}$
throughout the paper.}
\bibitem{note3}  {Below the
factor $\frac{\pi}{2}g$ is absorbed by the flux $\frac{\pi}{2}g
\phi^2\rightarrow \phi^2$.}
\bibitem{note4}
{It is always
possible by a proper choice of parameters (e.g. the cross-section in
quasi-1d case) to implement this limit while
keeping the conductance $g$ fixed.}
\bibitem{note5}
{Equation (\ref{SE}) is valid in all cases where $g_{d}>>1$.
Therefore it is valid in the metallic
regime in all dimensions and at the critical point in $d=2+\epsilon$.
However in order for the equation to be written in this way one should consider
a renormalized $g_{d}= a(d)g$. For the relevant dimensionality we have:
$g_{1}=2\pi g$, $g_{2}=2\pi^2
g$, and $g_{3}=4\pi^2 g$, where $g=D/(L^2 \Delta)$.}
\bibitem{mistake}  There was a minor mistake in calculating the Jacobian in
Ref.~[\onlinecite{FyodMir95}]. The correct answer is given by Eq.(\ref{ja})
\bibitem{note6} Equation (\ref{sfd}) is known to be valid for $d= 2+\epsilon$
since the work of Refs.~[\onlinecite{b10,b11}].
\bibitem{note7} For $\beta=1$ the relation $\mu = 2\beta g_{2}$ follows 
immediately from Eq.(\ref{PL}).
For $\beta=2$ it is analogous to the result of Ref.[\onlinecite{b14}]
for the problem of current relaxation. Here we are assuming that it
remains valid for $\beta=4$ as well.
\end{thebibliography}
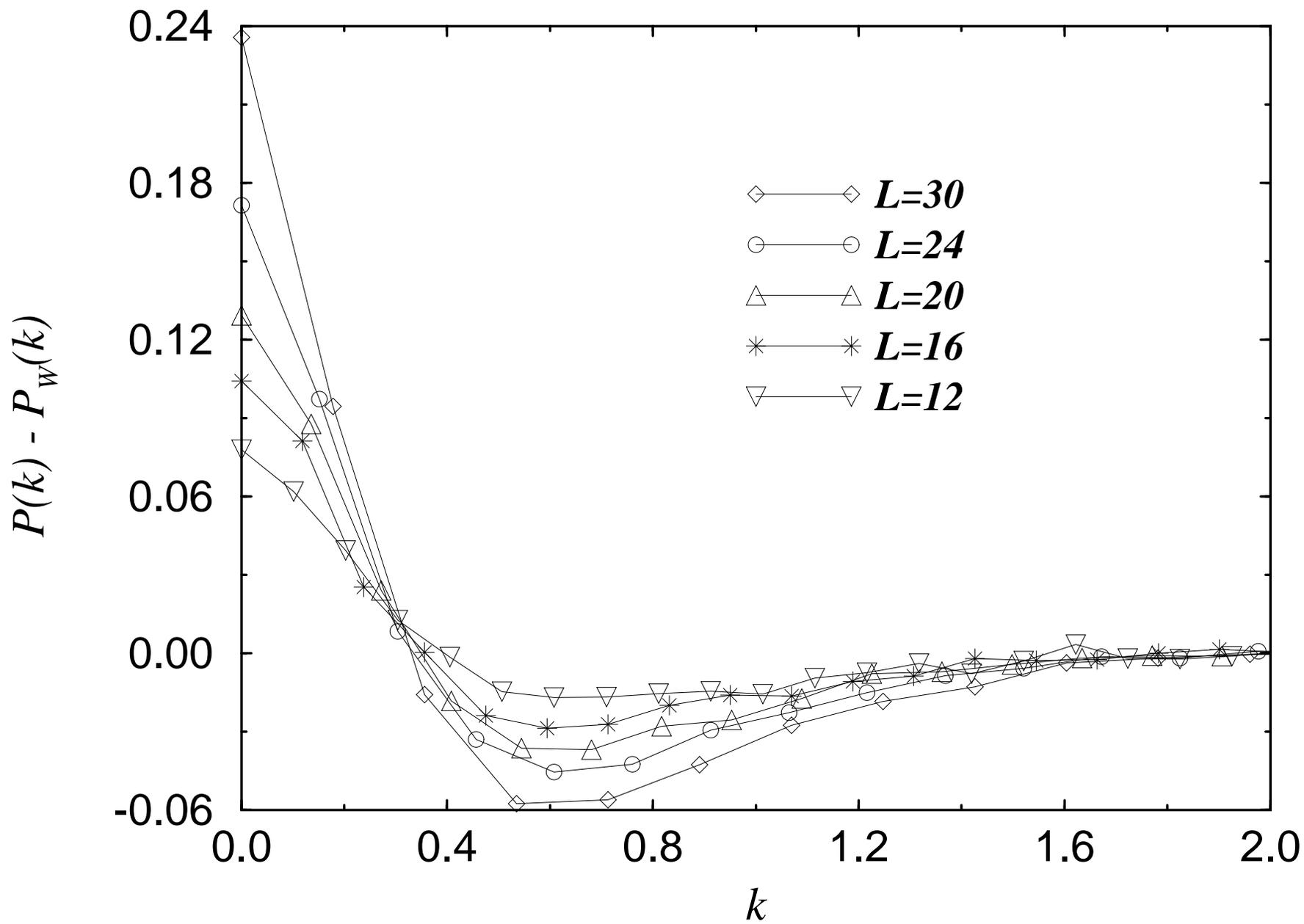
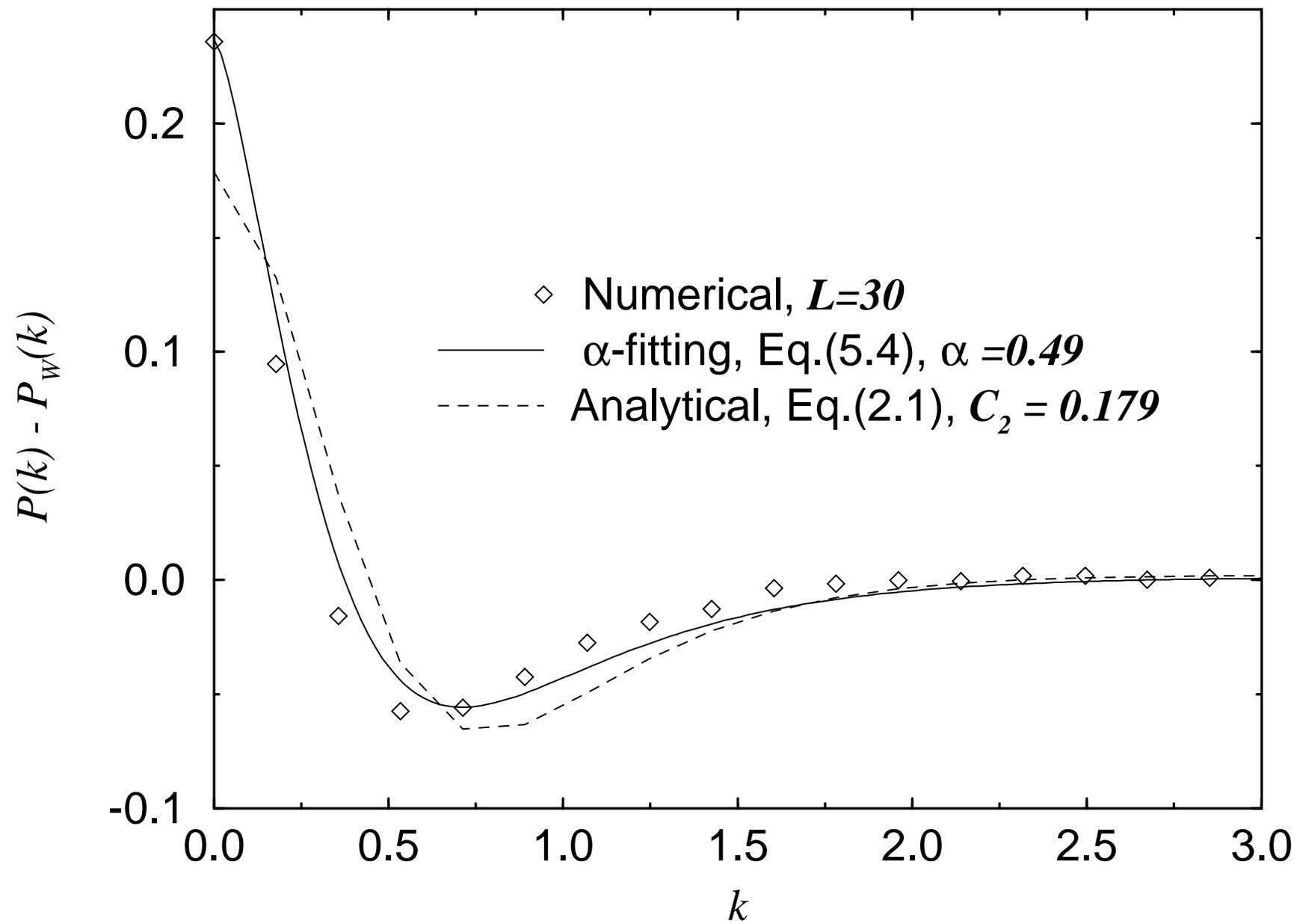
\end{document}